\documentclass[iop]{emulateapj}

\usepackage{hyperref}
\usepackage{color,hyperref}
\definecolor{linkcolor}{rgb}{0,0,0.5}
\definecolor{notecolor}{rgb}{0.8,0,0}
\hypersetup{colorlinks=true, linkcolor=linkcolor, citecolor=linkcolor, 
  filecolor=linkcolor, urlcolor=linkcolor}

\usepackage{amsmath}

\newcommand\xf{x_{\rm{f}}}

\newcommand\cm{\;{\rm cm}}
\newcommand\pc{\;{\rm pc}}

\newcommand\Myr{\;{\rm Myr}}

\newcommand\pcc{\;{\rm cm}^{-3}}

\newcommand\tgrow{t_{\rm gr}}
\newcommand\jx{j_{x0}}
\newcommand\Mx{M_{x0}}

\newcommand\FLen{\ell_{\rm{F}}}
\newcommand\DL{\ell_{\rm{D}}}
\newcommand\DLone{\ell_{\rm{D},1}}
\newcommand\FLone{\ell_{\rm{F},1}}
\newcommand\FLtwo{\ell_{\rm{F},2}}
\newcommand\lmax{\lambda_{\rm{max}}}
\newcommand\Peq{P_{\rm eq}}
\newcommand\Psat{P_{\rm{sat}}}
\newcommand\Pmin{P_{\rm{min}}}
\newcommand\Pmax{P_{\rm{max}}}

\newcommand\kms{\;{\rm km}\,{\rm s}^{-1}}
\newcommand\cms{\;{\rm cm}\,{\rm s}^{-1}}
\newcommand\ms{\;{\rm m}\,{\rm s}^{-1}}
\newcommand\ergs{\;{\rm erg}\,{\rm s}^{-1}}

\newcommand\Kel{\;{\rm K}}
\newcommand\Tpt{T^{\prime}}
\newcommand\punit{\pcc\,{\rm K}}
\newcommand\condunit{\;{\rm erg}\,{\rm cm}^{-1}\,{\rm s}^{-1}\,{\rm K}^{-1}}
\newcommand\kB{{\,k_{\rm B}}}
\newcommand\xim{\xi_-}
\newcommand\xip{\xi_+}
\newcommand\Lf{L_{\rm f}}

\newcommand\betam{\beta_{t-}}
\newcommand\betap{\beta_{t+}}

\newcommand{\p}{\partial}
\newcommand{\pr}{\prime}
\newcommand\simgt{\lower.5ex\hbox{$\; \buildrel > \over \sim \;$}}
\newcommand\simlt{\lower.5ex\hbox{$\; \buildrel < \over \sim \;$}}

\shorttitle{Instability of Evaporation Fronts} %
\shortauthors{Kim \& Kim}

\begin{document}

\title{Instability of Evaporation Fronts in the Interstellar Medium}

\author{Jeong-Gyu Kim and Woong-Tae Kim}

\affil{Center for the Exploration of the Origin of the Universe
  (CEOU), Astronomy Program, Department of Physics \& Astronomy,\\
  Seoul National University, Seoul 151-742, Republic of Korea}
\email{jgkim@astro.snu.ac.kr, wkim@astro.snu.ac.kr}
\slugcomment{Draft version}

\begin{abstract}
  The neutral component of the interstellar medium is segregated into
  the cold neutral medium (CNM) and warm neutral medium (WNM) as a
  result of thermal instability. It was found that a plane-parallel
  CNM--WNM evaporation interface, across which the CNM undergoes
  thermal expansion, is linearly unstable to corrugational
  disturbances, in complete analogy with the Darrieus-Landau
  instability (DLI) of terrestrial flames. We perform a full linear
  stability analysis as well as nonlinear hydrodynamic simulations of
  the DLI of such evaporation fronts in the presence of thermal
  conduction. We find that the DLI is suppressed at short length
  scales by conduction. The length and time scales of the fastest
  growing mode are inversely proportional to the evaporation flow
  speed of the CNM and its square, respectively.  In the nonlinear
  stage, the DLI saturates to a steady state where the front deforms
  to a finger-like shape protruding toward the WNM, without generating
  turbulence. The evaporation rate at nonlinear saturation is larger
  than the initial plane-parallel value by a factor of $\sim 2.4$ when
  the equilibrium thermal pressure is $1800\kB\punit$.  The degrees of
  front deformation and evaporation-rate enhancement at nonlinear
  saturation are determined primarily by the density ratio between the
  CNM and WNM.  We demonstrate that the Field length in the thermally
  unstable medium should be resolved by at least four grid points to
  obtain reliable numerical outcomes involving thermal instability.
\end{abstract}


\section{Introduction}\label{s:intro}

The interstellar medium (ISM) is inhomogeneous, consisting of multiple
components with a wide range of densities and temperatures.  In a
simple description of the two-phase model \citep{spi58,fie69}, a
diffuse gas suffers from thermal instability (TI) and segregates into
a cold neural medium (CNM) with temperature $T\sim10^2\Kel$ and a warm
neutral medium (WNM) with $T\sim 10^4\Kel$ (\citealt{fie65}; see also
\citealt{mee96} and \citealt{cox05} for reviews). Strong radiative and
mechanical heating by supernova explosions produces a hot third phase
that fills most of the volume in galaxies \citep{cox74,mo77}. The ISM
is highly responsive and thus changes its phase readily depending on
environmental conditions. For example, local compression and radiative
cooling turn a hot gas to a WNM and then to a CNM, while the reverse
phase transitions can occur due to expansion and heating (e.g.,
\citealt{mo07}). Despite pervasive presence of supersonic turbulence
in the ISM (e.g., \citealt{mac04,hei04}), pressure equilibrium among
different phases roughly holds as long as the characteristic time
between shocks is longer than the cooling time \citep{wol03}.

Phase transitions of the ISM usually involve interfaces or thermal
fronts between different phases (e.g., \citealt{sto11}). The thickness
of thermal fronts is of the order of the ``Field length''
\citep{fie65}, across which conductive heat flux balances the
radiative heating and cooling (see also \citealt{beg90}). In the case
of diffuse ISM, thermal fronts are occupied by gas in the TI-unstable
temperature range whose mass fraction is non-negligible compared to
the CNM and WNM (e.g.,
\citealt{pio05,pio07,hen07,kim08,kim10}). Thermal fronts are further
termed evaporation fronts when a colder component becomes hotter as it
moves across them, or condensation fronts in the opposite
situations. In pioneering studies, \citet{zel69} and \citet{pen70}
independently examined steady-state structure of planar thermal
fronts. They found that there exists the saturation pressure $\Psat$
at which a front experiences no net cooling, and that the equilibrium
thermal pressure $\Peq$ determines the type of thermal fronts such
that the fronts are static (i.e., no gas motion across them) when
$\Peq=\Psat$, while $\Peq > \Psat$ for condensation fronts and $\Peq <
\Psat$ for evaporation fronts due to excessive cooling or heating (see
also \citealt{ino06,iwa12}). \citet{sto10} considered magnetized
thermal fronts and showed that magnetic fields are distributed almost
uniformly due to efficient ambipolar diffusion, making the temperature
profile almost the same as in the unmagnetized cases.

As has often been noted, the mathematical problem of determining the
structure of thermal fronts in the ISM is identical to that of
terrestrial flames in combustion theory. In the case of evaporation
fronts, for example, an upstream CNM changing to a downstream WNM due
to radiative heating is analogous exactly to upstream unburnt gas
transforming to downstream burnt ash with chemical reactions as a
heating source. It has long been well recognized in combustion theory
that planar flame fronts are unconditionally unstable to front
distortions owing to thermal expansion across them (e.g.,
\citealt{wil85,zel85,lib94,law06,sea09}; see \citealt{byc00} for an
in-depth review). This corrugational instability is usually referred
to as the Darrieus-Landau instability (DLI) after the original studies
of \citet{dar38} and \citet{lan44}. When the flow is assumed
incompressible and the flame front is taken infinitesimally thin
(i.e., ignoring the effect of thermal conduction), the growth rate
$\Omega_0$ of the DLI is given by
\begin{equation}\label{e:Omega0}
  \Omega_0 = k v_{x1}\dfrac{\mu}{1 + \mu}\left(\sqrt{1 + \mu -
      \mu^{-1}} -1\right),
\end{equation}
where $k$ is the wavenumber of perturbations transverse to the flow
direction, $v_{x1}$ is the velocity of the upstream unburnt gas with
respect to the front, and $\mu\; (>1)$ is the expansion factor defined
as the ratio of unburnt to burnt gas densities
\citep[e.g.,][]{wil85,lan87}. \citet{lib94} studied the linear
stability of a flame front with finite thickness, showing that thermal
conduction stabilizes short-wavelength perturbations.  \citet{tra97}
further considered the effect of flow compressibility and found that
the maximum growth rate increases with the flow Mach number.

The DLI has been of special interest in the study of explosive
nucleosynthesis occurring in Type Ia supernova flames as it is
considered as one of the candidate mechanisms that may trigger the
deflagration-to-detonation transition in thermonuclear burning
\citep{nie95,rop03,dur03,bel04a}. In particular, \citet{bel04a}
carried out high-resolution numerical simulations of the DLI of C/O
thermonuclear flames with $\mu\sim1.4-1.7$ by including the effect of
finite flame thickness, corresponding to the late stages of a Type Ia
supernova event. They found that the DLI in the linear stage
accelerates the flames by increasing their surface area, and saturates
in the nonlinear stage by forming round cusps in the flames. In their
models, the maximum enhancement in the flame speeds is only a few
percents, about an order of magnitude smaller than the results of
\citet{rop03} that treated the flames as being infinitesimally
thin. \citet{bel04a} also showed that it is important to resolve flame
fronts by at least 5--10 zones to obtain reliable simulation outcomes
(see also \citealt{bel04b}).

\citet{ino06} noted that the DLI arises in evaporation fronts in the
diffuse ISM, as well. In addition to finding the dependence of the
evaporation or condensation rate on the equilibrium pressure, they
obtained an approximate dispersion relation of the DLI by considering
long- and short-wavelength perturbations separately.  They showed that
their results recover Equation \eqref{e:Omega0} in the long wavelength
limit, and that the DLI is suppressed on small scales by thermal
conduction. Noting that the growth time and preferred length scales of
the instability can be comparable to the cooling time and the Field
length, respectively, they proposed that the DLI can be a driving
mechanism for ISM turbulence. \citet{sto10} analyzed linear stability
of evaporation and condensation fronts with magnetic fields embedded
orthogonal to the fronts, showing that evaporation fronts are
stabilized by magnetic fields when the flows are sub-Alfv\'{e}nic, a
most likely situation in the ISM.

Despite these efforts, nonlinear outcomes of the DLI of evaporation
fronts have yet to be explored in order to assess its dynamical
consequences to the ISM.  Most of all, it is not clear whether the DLI
will drive turbulence, as envisioned by \citet{ino06}, or simply
saturate nonlinearly as in the case of thermonuclear flames. The
expansion factor for evaporation fronts in the ISM is typically
$\mu\sim 40$--$200$ \citep{wol03}, more than an order of magnitude
larger than those in the chemical or thermonuclear flames, so that it
is interesting to study the effect of $\mu$ on the changes in the
evaporation rate and front shapes in the nonlinear stage.  In
addition, a full linear stability analysis of the DLI, applicable to a
CNM-WNM interface in the diffuse ISM, that properly takes allowance
for finite front thickness is still lacking.  We therefore in this
paper investigate both linear stability and nonlinear evolution of
evaporating fronts in the ISM by including the effect of thermal
conduction. For the linear stability analysis, we follow the
eigenvalue approach of \citet{lib94} and find numerical dispersion
relations for general $k$. We run hydrodynamic simulations to study
nonlinear development of the DLI. We also study the effect of $\mu$ on
the nonlinear state of the DLI by employing a modified form of the
heating function, and demonstrate that sufficient resolution is
required to resolve interfaces between the CNM and WNM.

The rest of this paper is organized as follows. In Section
\ref{s:theory}, we introduce basic equations of hydrodynamics and
calculate the structure of thermal fronts in a steady state.  In
Section \ref{s:lin}, we present the result of full linear stability
analyses and the scaling relations for the most unstable modes. In
Section \ref{s:1D}, we present the requirement of numerical resolution
to resolve the thermal interfaces between CNM and WNM.  The results of
two-dimensional simulations of the DLI using both single-mode and
multi-mode perturbations as well as the effects of varying expansion
factor are presented in Section \ref{s:2D}. We summarize and discuss
our main results in Section \ref{s:sum}.

\section{Steady Fronts}\label{s:theory}

\subsection{Basic Equations}

In this paper we consider gas flows across an evaporating front
between the CNM and WNM of the ISM, and study stability of the front
against distortional perturbations in the presence of thermal
conduction. We do not consider the effect of magnetic fields and
gaseous self-gravity in the present work.  The governing equations of
ideal hydrodynamics read
\begin{equation}\label{e:cont}
  \frac{\p \rho}{\p t} + \nabla \cdot (\rho \mathbf{v}) = 0\,,
\end{equation}
\begin{equation}\label{e:mom}
  \frac{\p \rho \mathbf{v}}{\p t} + \nabla \cdot \left( \rho
    \mathbf{v}\mathbf{v} + P\mathbf{I} \right) = 0\,,
\end{equation}
\begin{equation}\label{e:e}
  \frac{\p E}{\p t} + \nabla \cdot \left( (E + P)\mathbf{v}
                   - \kappa \nabla T
  \right) = -\rho\mathcal{L}\,,
\end{equation}
where $\rho$, $\mathbf{v}$, and $P$ are the gas density, velocity, and
pressure, respectively, $E = P/(\gamma - 1) + \rho v^2/2$ is the total
energy density with the adiabatic index $\gamma=5/3$, $\kappa$ is the
thermal conductivity, and $\rho\mathcal{L}$ is the net radiative
cooling rate per unit volume.  For the thermal pressure, we take an
ideal gas law
\begin{equation}\label{e:ideal}
  P = \frac{\rho \kB T}{\overline{m}}\,,
\end{equation}
where $\kB$ is the Boltzmann constant and $\overline{m} = 1.37
m_{\rm{H}}$ denotes the mean mass per hydrogen atom, corresponding to
the solar abundances. In this work, we ignore the effect of viscosity
since it has little influence on the dynamics of thermal fronts
\citep{pel82}.

The net volumetric heat-loss rate is given by
\begin{equation}\label{e:nCool}
  \rho\mathcal{L} = n^2\Lambda(T) - n \Gamma\,,
\end{equation}
with $n=\rho/\overline{m}$ being the number density of hydrogen. In
the diffuse ISM, the heating rate $\Gamma$ is primarily by the
photoelectric effect on small dust grains and polycyclic aromatic
hydrocarbons by FUV radiation \citep{bak94}, for which we take the
solar-neighborhood value
\begin{equation}\label{e:heatfn2}
  \Gamma =\Gamma_0= 2.0 \times 10^{-26} \ergs \,,
\end{equation}
adopted by \citet{koy02}.  The radiative cooling function $\Lambda$ is
dominated by the fine-structure lines of \ion{C}{2} at low $T$ and
Ly$\alpha$ emissions at high $T$, for which we adopt the fitting
formula suggested by \citet{koy02}:
\begin{equation}\label{e:heatfn3}
  \frac{\Lambda(T)}{\Gamma_0} = 10^7 \exp \left(\dfrac{-1.184 \times
      10^5}{T + 1000}\right) + 0.014 \sqrt{T}
  \exp\left(\dfrac{-92}{T}\right) \cm^3 \,,
\end{equation}
where $T$ is in units of degrees Kelvin (see also \citealt{vaz07}).

\begin{figure}
  \epsscale{1.2} \plotone{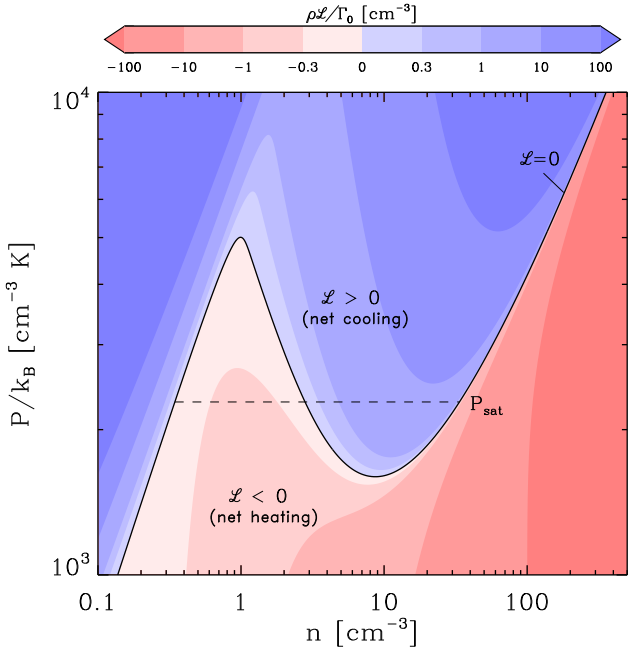}
  \caption{Thermal equilibrium curve with $\mathcal{L}=0$ (solid line)
    in the density -- pressure plane overlaid over shaded contours
    of our adopted heat-loss function. The regions above and below
    the curve correspond to gas with net cooling and heating,
    respectively. Two-phase equilibria for the coexistence of the CNM
    and WNM are possible for the equilibrium pressure in the range
    between $\Pmin/\kB=1597\punit$ and $\Pmax/\kB=5005\punit$, and a
    static equilibrium is attained at the saturation pressure
    $\Psat/\kB = 2282 \punit$, marked as the horizontal dashed line.
  }\label{f:np}
\end{figure}

Figure \ref{f:np} plots shaded contours of the heat-loss function in
the $n$--$P$ plane together with the locus of points where cooling
balances heating (i.e., $\mathcal{L}=0$) as the solid curve. The
regions above (below) the equilibrium curve are dominated by cooling
(heating). Thermally unstable gas is located on the portion of the
curve with $d\log P/d\log n < 0$. The gas with $n> 8.7\pcc$ and
$T<185\Kel$ along the curve is referred to as the CNM, while the WNM
has $n<1\pcc$ and $T>5001\Kel$. The CNM and WNM can coexist only when
the pressure is in the range $\Pmin < P < \Pmax$, where
$\Pmin/\kB=1597\punit$ and $\Pmax/\kB=5005\punit$ are the minimum and
maximum pressures for two-phase equilibria.

In the neutral ISM, thermal conduction is mostly due to collisions of
hydrogen atoms.  The corresponding conductivity depends on the
temperature as $\kappa = 2.5 \times 10^3 \sqrt{T} \condunit$
\citep{par53,spi62}. As we shall show below, it is important to
resolve the Field length in numerical simulations in order to obtain
reliable results. While it is desirable to use the above form of
thermal conductivity, we found that this requires extremely high
resolution to resolve a transition layer close to the CNM.  Throughout
this work, therefore, we take a constant value $\kappa_0 = 10^5
\condunit$, corresponding to $T = 1600 \Kel$. While using a constant
conductivity slightly changes the thickness of a transition layer
between the CNM and WNM, it does not alter the essential physics
involved in the DLI.

\subsection{Front Structure}\label{s:equil}

We seek for one-dimensional, steady-state solutions of Equations
\eqref{e:cont}--\eqref{e:e} that allow a phase transition between the
CNM and WNM.  We work in a frame in which the transition front is
stationary. Following \citet{ino06} and \citet{iwa12}, we take a
convention that the CNM and WNM are located at the left- and
right-hand sides of the $x$-axis, respectively. Equations
\eqref{e:cont}--\eqref{e:e} are then simplified to
\begin{equation}\label{e:j}
  \jx \equiv \rho v_x = \text{constant}\,,
\end{equation}
\begin{equation}\label{e:M}
  \Mx \equiv P + \rho v_x^2 = \text{constant}\,,
\end{equation}
\begin{equation}\label{e:stn}
  \kappa\frac{d^2T}{d x^2}  = \jx c_P \frac{d T}{dx} + \rho \mathcal{L}(T)\,.
\end{equation}
where $\jx$ and $\Mx$ denote the mass flux and the momentum flux,
respectively, and $c_P = \gamma(\gamma-1)^{-1} k_B/\overline{m}$ is
the specific heat at constant pressure. Note that in deriving Equation
\eqref{e:stn} we make the isobaric approximation $P \approx
\text{constant}$, the validity of which will be justified a
posteriori.

Equation (\ref{e:stn}) suggests that there are two characteristic
length scales related to a transition layer:
\begin{align}\label{e:len}
  \DL = \dfrac{\kappa/\rho c_P}{v_{x}},\;\;\;\;\;\text{and}\;\;\;\;
  \FLen = \sqrt{\frac{\kappa T}{n \Lambda^2}}\,.
\end{align}
The first one is the heat diffusion length occurring over the
advection time scale ($\DL/v_x$), while the second one is the Field
length over which the conductive heat transport balances the cooling
\citep{beg90}. The latter also corresponds to the maximum wavelength
of TI in the presence of thermal conduction \citep{fie65}. For
thermal fronts in the ISM we consider here, the advection term in
Equation \eqref{e:stn} is much smaller than the heating and
conduction terms (i.e., $\DL \gg \FLen$), so that $\FLen$ naturally
corresponds to the thickness of transition layers in the
ISM.\footnote{This is unlike in
  terrestrial flame fronts where $\DL\ll\FLen$, so that the front
  thickness is determined primarily by $\DL$ (e.g., \citealt{zel85}).}
Note that $\FLen/\DL = t_{\rm{cool}} / t_{\rm{flow}}$, where
$t_{\rm{cool}} = \gamma(\gamma-1)^{-1} P / n^2\Lambda$ is the cooling
time and $t_{\rm{flow}} = \FLen /v_{x}$ is the time it takes a fluid
element to pass through the front.

Equation \eqref{e:stn} can be integrated numerically subject to the
conditions
\begin{equation}\label{e:bd}
  T\bigl\vert_{-\infty} =     T_1\,,\;\;\;\;\;
  T\bigl\vert_{+\infty} =     T_2\,,\;\;\;\;\;
  \dfrac{d T}{dx}\Bigl\vert_{\pm \infty} = 0\,.
\end{equation}
Here and hereafter, the subscripts ``1'' and ``2'' indicate the
physical quantities of the CNM and WNM very far away from the front,
respectively, at an equilibrium pressure $\Peq$.  Thus, finding $T(x)$
constitutes an eigenvalue problem with eigenvalue $\jx$. For given
$\Peq$, we take a trial value of $\jx$ and integrate Equation
\eqref{e:stn} from $x=\pm\infty$ toward a midpoint where $T$'s from
both sides match with each other. We check if $dT/dx$ from each side
is the same at the midpoint as well, and vary $\jx$ iteratively until
the smoothly connecting solutions are obtained. By repeating the
procedures, one can find $\jx$ as a function of $\Peq$ for steady
equilibria.

\begin{figure}
  \epsscale{1.2} \plotone{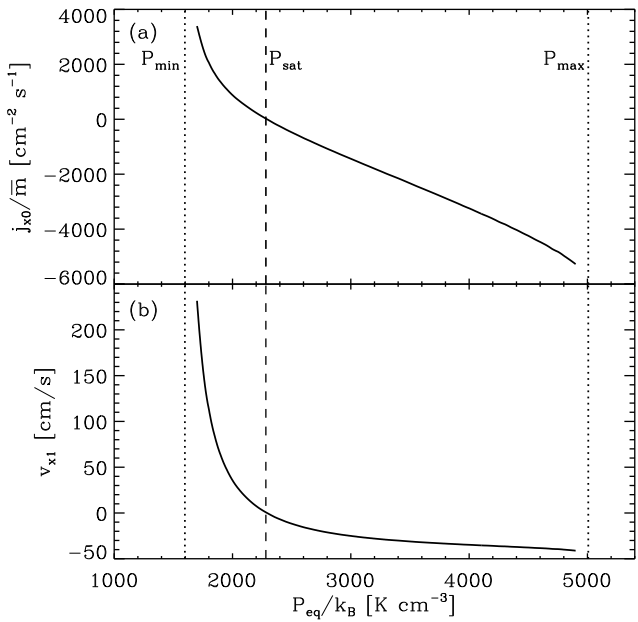}
  \caption{(a) Equilibrium mass flux $\jx$ and (b) the inflow speed
    $v_{x1}$ of the CNM at $x=-\infty$ for steady fronts between
    the CNM and WNM as functions of $\Peq$. The front is static
    ($\jx=0$) at $\Peq = \Psat$ indicated by the vertical dashed line,
    and corresponds to an evaporation front when $\Peq<\Psat$ and a
    condensation front when $\Peq>\Psat$. The vertical dotted lines
    mark $\Pmax$ and $\Pmin$.}\label{f:vx1}
\end{figure}

Figure \ref{f:vx1} plots (a) $\jx/\overline{m}$ and (b) the
equilibrium CNM velocity $v_{x1}=\jx /\rho_1$ at $x=-\infty$ as
functions of $\Peq$. The front is static (i.e., $\jx=v_{x1}=0$) at the
saturation pressure $\Psat /\kB = 2282 \punit$ marked by the dashed
line in Figures \ref{f:np} and \ref{f:vx1}. Both $|\jx|$ and
$|v_{x1}|$ increase as $\Peq$ departs from $\Psat$. Since $T_2\gg
T_1$, Equation \eqref{e:stn} is further integrated to
\begin{equation}\label{e:jQ}
\jx \approx \mathcal {Q} / (c_P T_2)\,,
\end{equation}
where $\mathcal{Q} \equiv -\int_{-\infty}^\infty \rho \mathcal{L} dx$,
indicating that $\mathcal{Q}=0$ for static fronts \citep{zel69}. When
$\Peq < \Psat$, the equilibrium densities are smaller than the static
cases and the gas is thus dominated by heating with
$\mathcal{Q}>0$. In this case, the transition layer corresponds to an
evaporation front since the CNM moves in the positive $x$-direction
and undergoes thermal expansion to turn to the WNM downstream. When
$\Peq>\Psat$, on the other hand, the radiative cooling dominates to
have $\mathcal{Q}<0$ and $v_{x1}<0$, so that the WNM moves in the
negative $x$-direction to change to the CNM after passing through a
condensation front. The eigenvalue $\jx$ is identical to the
evaporation or condensation rate of the gas per unit area across the
front. Figure \ref{f:vx1} shows that for evaporation fronts, the
inflowing CNM velocity at far upstream is larger for smaller $\Peq$,
but is limited to below $2.5 \ms$. The corresponding WNM velocity at
far downstream amounts to $v_{x2}=(\rho_1/\rho_2)v_{x1} < 137
\ms$. Since the associated Mach number is less than $0.016$, one can
ignore the $\rho v_x^2$ term in Equation \eqref{e:M} to obtain
$P\approx \Mx =\text{constant}$. This proves the validity of the
isobaric approximation for steady fronts.

\begin{figure}
  \epsscale{1.2} \plotone{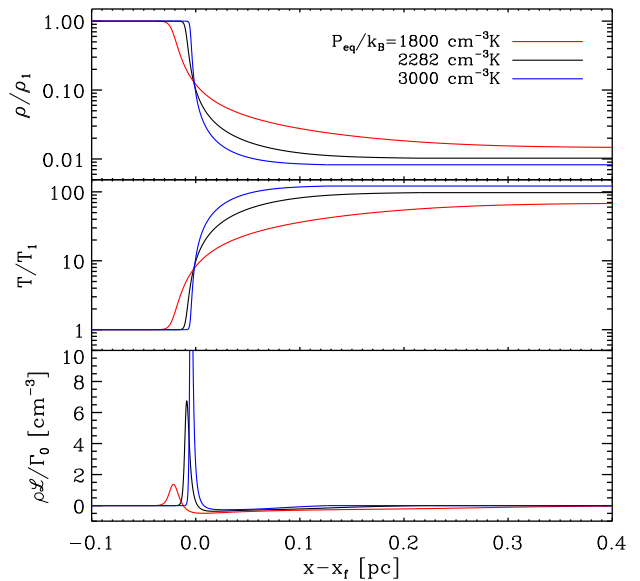}
  \caption{Distributions of density (top), temperature (middle), and
    local net cooling rate (bottom) for an evaporation front
    with $\Peq/\kB=1800\punit$ (red), a static front with
    $\Peq/\kB = 2282\punit$ (black), and a condensation front with
    $\Peq/\kB = 3000\punit$ (blue).  Cooling regions are highly localized
    near to the CNM side of the transition layer where $\rho$ and $T$
    vary steeply.}\label{f:t0}
\end{figure}

\begin{deluxetable}{ccccccccc}
\tabletypesize{\footnotesize} \tablewidth{0pt}
\tablecaption{Properties of Evaporation Fronts} \tablehead{
\colhead{$\Peq/\kB$} & %
\colhead{$\mu$}      & %
\colhead{$v_{x1}$}   & %
\colhead{$\DLone$}   & %
\colhead{$\FLone$}   & %
\colhead{$\FLtwo$}   & %
\colhead{$\Lf$}      & %
\colhead{$\lmax$}    & %
\colhead{$\tgrow$}       \\
\colhead{(1)}  & \colhead{(2)} & \colhead{(3)} & \colhead{(4)} &
\colhead{(5)}  & \colhead{(6)} & \colhead{(7)} & \colhead{(8)} &
\colhead{(9)}}
\startdata
1700 &  59.1 &  231.7 &  0.028 & 0.0020 & 0.12 & 0.43 & 0.798 & 21.0  \\
1800 &  68.9 &  113.6 &  0.045 & 0.0017 & 0.12 & 0.29 &  1.05 & 42.8  \\
1900 &  76.5 &   63.1 &  0.069 & 0.0015 & 0.11 & 0.22 &  1.43 & 91.1  \\
2000 &  83.0 &   35.8 &  0.106 & 0.0013 & 0.11 & 0.19 &  2.10 &  219  \\
2100 &  88.7 &   18.8 &  0.179 & 0.0011 & 0.11 & 0.16 &  3.58 &  674
\enddata
\tablecomments{Column 1: equilibrium pressure (cm$^{-3}$ K). Column 2:
  $\mu=\rho_1/\rho_2$ is the expansion factor. Column 3: the inflow velocity of
  the CNM at $x=-\infty$ (cm s$^{-1}$). Columns 4--7: the diffusion length in
  the CNM, Field length in the CNM, Field length in the WNM, and front
  thickness, respectively (pc). Columns 8--9: the wavelength (pc) and the growth
  time (Myr) of the most unstable mode of the DLI, respectively.}\label{t:front}
\end{deluxetable}

Figure \ref{f:t0} displays the exemplary distributions of (a) gas
density, (b) temperature, and (c) net heat-loss function for the
evaporation (red), static (black), and condensation (blue) fronts at
the equilibrium pressure $\Peq /\kB = 1800$, $2282$, $3000 \punit$,
respectively. The various profiles are shifted such that the front
position, $\xf$, defined as the location where $T = \sqrt{T_1T_2}$
is the same for all cases. Note that for evaporation/condensation
fronts the temperature profile is identical to the velocity profile
under the isobaric approximation (i.e., $T/T_1 = v_x/v_{x1}$). For
the evaporation front, the expansion factor
$\mu=\rho_1/\rho_2=T_2/T_1$ is $68.9$ and the Field lengths are
$\FLone = 1.7\times 10^{-3} \pc$ in the CNM and $\FLtwo=0.12 \pc$ in
the WNM. For all fronts, temperature and density vary relatively
slowly with $x$ in the heating region where $\rho\mathcal{L} <0$,
while the region of excessive cooling is highly localized spatially
with steeply varying $\rho$ and $T$. The front thickness $\Lf$
defined by the distance over which $T$ changes from $1.1T_1$ to
$0.9T_2$ is $\Lf=0.13\pc=1.31\FLtwo$ and $\Lf=0.078\pc=0.93\FLtwo$
in the evaporation and condensation fronts, respectively,
demonstrating that the front thickness is of the order of the Field
length in the WNM. Columns (2)--(7) of Table \ref{t:front} list
$\mu$, $v_{x1}$, $\DLone$, $\FLone$, $\FLtwo$, and $\Lf$ of
evaporation fronts for five selected values of $\Peq$. Accurate
modeling of thermal interfaces requires to resolve these rapid
changes in $T$ and $\rho\mathcal{L}$ in the cooling-dominated
region, as we will show in Section \ref{s:1D}.

\section{Linear Dispersion Relation}\label{s:lin}

\begin{figure}
  \epsscale{1.2} \plotone{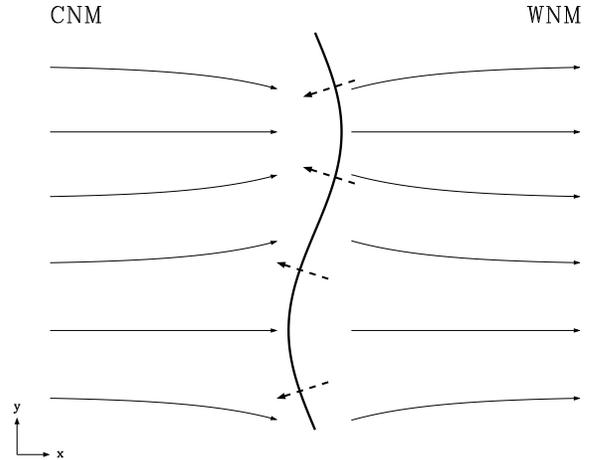}
  \caption{Schematic diagram showing directions of gas flows (solid
    arrows) near a distorted evaporation front (thick curve) that was
    originally parallel to the $y$-direction.  The CNM and WNM are
    located at the upstream and downstream sides, respectively.  The
    dashed arrows indicate the directions of heat flows via
    conduction. See text for details.}\label{f:diag}
\end{figure}

We explore the DLI of a steady evaporation front in the linear regime.
We begin by summarizing the physics behind the DLI in the long
wavelength limit: the reader is referred to \citet{wil85} and
\citet{zel85} for a more detailed explanation. Figure \ref{f:diag}
sketches a situation where a front is displaced sinusoidally (thick
solid line) along the $y$-direction, with a few solid arrows
representing the direction of gas motions near the front. The CNM and
WNM are located, respectively, at the left- and right-hand side of the
front that is approximated as a discontinuous surface.

After passing the evaporating front, the flow is refracted
\emph{toward} the normal to the front due to expansion, which is in
stark contrast to the cases of shock fronts where a postshock flow is
refracted \emph{away} from the front normal. As a result, streamlines
at the parts convex (concave) toward the WNM diverge (converge) behind
the front, decreasing (increasing) pressure there.  These pressure
changes behind the front affect the streamlines ahead of the front in
such a way that the upstream gas is directed toward (away from) the
convex (concave) parts. This in turn makes the mass flux at the convex
(concave) parts larger (smaller) than the local evaporation rate per
unit area which should be constant over time and space (see
\citealt{lib94} for a rigorous proof).  As the amount of gas
approaching is more (less) than can be afforded at the convex
(concave) part, the front there should advance further downstream
(upstream). This causes the front to bend in a runaway fashion,
indicating an instability.\footnote{On the other hand, a condensation
  front is stable to distortional perturbations, with $\text{Re}(\Omega_0) <0$
  when $\mu<1$ from Equation \eqref{e:Omega0}.}

The above argument is valid as long as the wavelengths of
perturbations are much longer than the front width. The growth rate
$\Omega_0$ given in Equation \eqref{e:Omega0} implies that modes with
smaller wavelength grow faster. However, small-scale modes would be
stabilized due to heat conduction by the following manner.  As
indicated by the dashed arrows in Figure \ref{f:diag}, excess heat in
the WNM can be easily transferred via conduction to the CNM ahead of
the convex parts of the front. The enhanced heating would speed up the
evaporation rate, compensating at least partly for the increased mass
flux there and thereby reducing a need for the front to advance
further. If this conduction-mediated evaporation rate exceeds the
increased mass flux, the front would be drawn back to the original
position and the DLI would be completely suppressed.  The effect of
conduction is important when the perturbation wavelength is comparable
to or less than the front thickness.

In Appendix \ref{a:lin}, we present the detailed procedure for finding
the linear dispersion relations of the DLI of an evaporation front in
the presence of conduction. Following the method of \citet{lib94}, we
describe the perturbations as a linear combination of incompressible,
vortex, and thermal modes in the far upstream and downstream sides
separately, and make them connect smoothly to each other at the front,
which allows us to obtain the growth rate $\Omega$ uniquely for given
wavenumber $k$ and $\mu$. We also show analytically that
$\Omega\rightarrow \Omega_0$ as $k\rightarrow0$.  While our method
requires cumbersome iterative integrations of the linearized
equations, it does not require to make long- or short-wavelength
approximations.

\begin{figure}
  \epsscale{1.2} \plotone{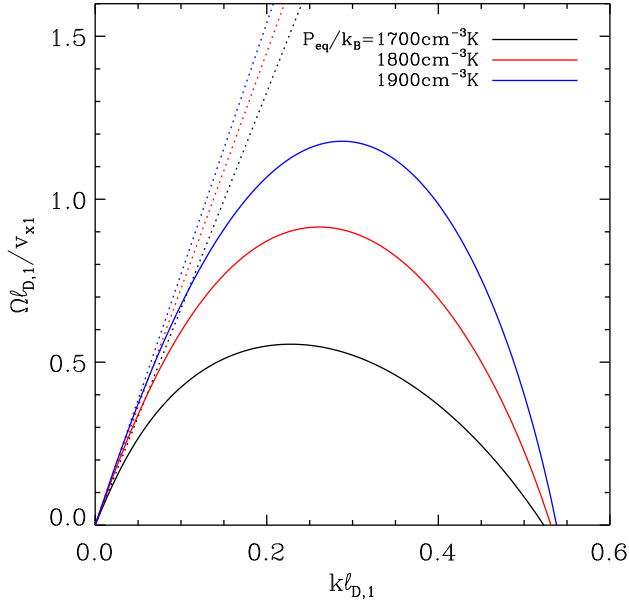}
  \caption{Dimensionless growth rate $\Omega \DLone/v_{x1}$ of the DLI
    for $\Peq/\kB=1700$, 1800, and $1900\punit$ as functions of the
    dimensionless wavenumber $k\DLone$.  Solid curves are the results
    of the full linear stability analysis, while dotted lines draw Equation
    \eqref{e:Omega0} with the corresponding $\mu$.}\label{f:gr}
\end{figure}

Figure \ref{f:gr} plots $\Omega$ against $k$ as solid lines for
$\Peq/\kB=1700$, $1800$, and $1900\punit$. The growth rate and
wavenumber are normalized using $v_{x1}$ and $\DLone$. The dotted
lines draw the corresponding $\Omega_0$, which are in good agreement
with $\Omega$ in the limit of $k\DLone\ll1$. The growth rate achieves
its peak value at $k_{\rm max}\DLone \sim 0.23$--$0.29$, is slightly
asymmetric with respect to $k_{\rm max}$, and becomes zero at $k_{\rm
  crit}\DLone \sim 0.52$--$0.54$. Columns (8) and (9) of Table
\ref{t:front} give $\lmax=2\pi/k_{\rm max}$ and the growth time
$\tgrow=1/\Omega_{\rm max}$ of the fastest growing mode in physical
units.  These are fitted approximately as
\begin{equation}\label{e:maxlen}
  \lmax  = 1.75 \mathcal{F}_\lambda \pc \left(\dfrac{v_{x1}}{1
      \ms}\right)^{-1}\left(\dfrac{n_1}{10
      \pcc}\right)^{-1}\left(\dfrac{\kappa}{\kappa_0} \right)\,,
\end{equation}
and
\begin{equation}\label{e:maxtime}
  \tgrow= 55.4 \mathcal{F}_t \Myr \left(\dfrac{v_{x1}}{1
      \ms}\right)^{-2}\left(\dfrac{n_1}{10 \pcc}\right)^{-1}
  \left(\dfrac{\kappa}{\kappa_0} \right)\,,
\end{equation}
where $\mathcal{F}_\lambda (\Peq)=1+ 30.3 (1 - \Peq/\Psat)^3$ and
$\mathcal{F}_t (\Peq)= 1+111(1-\Peq/\Psat)^3$ are the correction
factors for $\Peq$ in the range of $0.74\leq\Peq/\Psat\leq0.92$. The
dependence of $\lmax$ and $\tgrow$ on $v_{x1}$, $n_1$ and $\kappa$
follows simply from $\lmax\propto \DLone$ and $\tgrow\propto
\lmax/v_{x1}$. These fitting formulae are accurate within 7\%. The DLI
of evaporation fronts at higher equilibrium pressure takes longer time
to grow, owing to a smaller $v_{x1}$ in the background state.

\section{Numerical Simulations}\label{s:num}

To study nonlinear development of the DLI of an equilibrium
configuration found in the preceding section, we evolve the set of
Equations \eqref{e:cont}--\eqref{e:e} by using the \emph{Athena} code
\citep{sto08}. \emph{Athena} is a general-purpose Eulerian code for
magnetohydrodynamics based on high-order Godunov methods. Among the
various algorithms implemented in it, we use the CTU scheme for
directionally unsplit integration, the HLLC Riemann solver for flux
computation, and the piecewise linear method for spatial
reconstruction. The thermal conduction and heat-loss terms are solved
explicitly. In this section, we first address the issue of proper
resolution required to resolve an interface between the CNM and WNM,
and then present the numerical results for the DLI in the nonlinear
regime.

\subsection{Constraint on Spatial Resolution}\label{s:1D}

To check our implementation of the heating, cooling, and conduction
terms in the \emph{Athena} code, we have tested the code to the
growth of TI by running one-dimensional simulations. For this
purpose, we initially consider a static, thermally unstable medium
with $n = 2.80 \pcc$ and $T=814 \Kel$ in the domain with size
$L_x=20\pc$, and impose random perturbations to the pressure with
amplitudes of $0.1\%$. We employ the periodic boundary conditions at
both ends of the domain. We run various models with differing number
of grid points from $N_x=2^7$ to $2^{14}$. Table \ref{t:TI} gives
the results of these one-dimensional simulations.   As Column (3) of
Table \ref{t:TI} shows, all of our runs successfully reproduce,
within $\sim4\%$, the analytic growth rate $\tau_{\rm gr}=0.85
\Myr^{-1}$ of the most unstable mode in the linear regime,
consistent with the results of previous studies (e.g.,
\citealt{pio04,kim08,cho12}).

\begin{deluxetable}{ccccc}
\tabletypesize{\footnotesize} \tablewidth{0pt} \tablecaption{Results
of One-dimensional Simulations of TI} \tablehead{
\colhead{$N_x$} & %
\colhead{$\Delta x$}   & %
\colhead{$\tau_{\rm gr}$}  & %
\colhead{$\Psat/\kB$}   & %
\colhead{$\delta v_x$}   \\
\colhead{(1)}  & \colhead{(2)} & \colhead{(3)} & \colhead{(4)} &
\colhead{(5)}  } \startdata
  $2^7$   & $1.6 \times 10^{-1}$ & 0.87 & 1818 &  $3.6 \times 10^{1}$ \\
  $2^8$   & $7.8 \times 10^{-2}$ & 0.84 & 1813 &  $2.7 \times 10^{1}$ \\
  $2^9$   & $3.9 \times 10^{-2}$ & 0.84 & 1865 &  $1.9 \times 10^{1}$ \\
  $2^{10}$ & $2.0 \times 10^{-2}$ & 0.85 & 1989 &  $1.4 \times 10^{1}$ \\
  $2^{11}$ & $9.8 \times 10^{-3}$ & 0.84 & 2158 &  $6.6 \times 10^{0}$ \\
  $2^{12}$ & $4.9 \times 10^{-3}$ & 0.86 & 2267 &  $2.4 \times 10^{0}$ \\
  $2^{13}$ & $2.4 \times 10^{-3}$ & 0.85 & 2282 &  $4.7 \times 10^{-2}$ \\
  $2^{14}$ & $1.2 \times 10^{-3}$ & 0.85 & 2282 &  $1.2 \times 10^{-2}$
\enddata
\tablecomments{Column 1--3: number of zones, the zone spacing (pc),
and the numerical growth rate of TI ($\rm{Myr}^{-1}$). Column 4--5:
numerically found equilibrium pressure ($\rm{cm}^{-3}\Kel$) and the
velocity dispersion ($\rm{m}\;\rm{s}^{-1}$)
  averaged over $t= 200$--$500 \Myr$.}\label{t:TI}
\end{deluxetable}

\begin{figure*}
  \epsscale{1.} \plotone{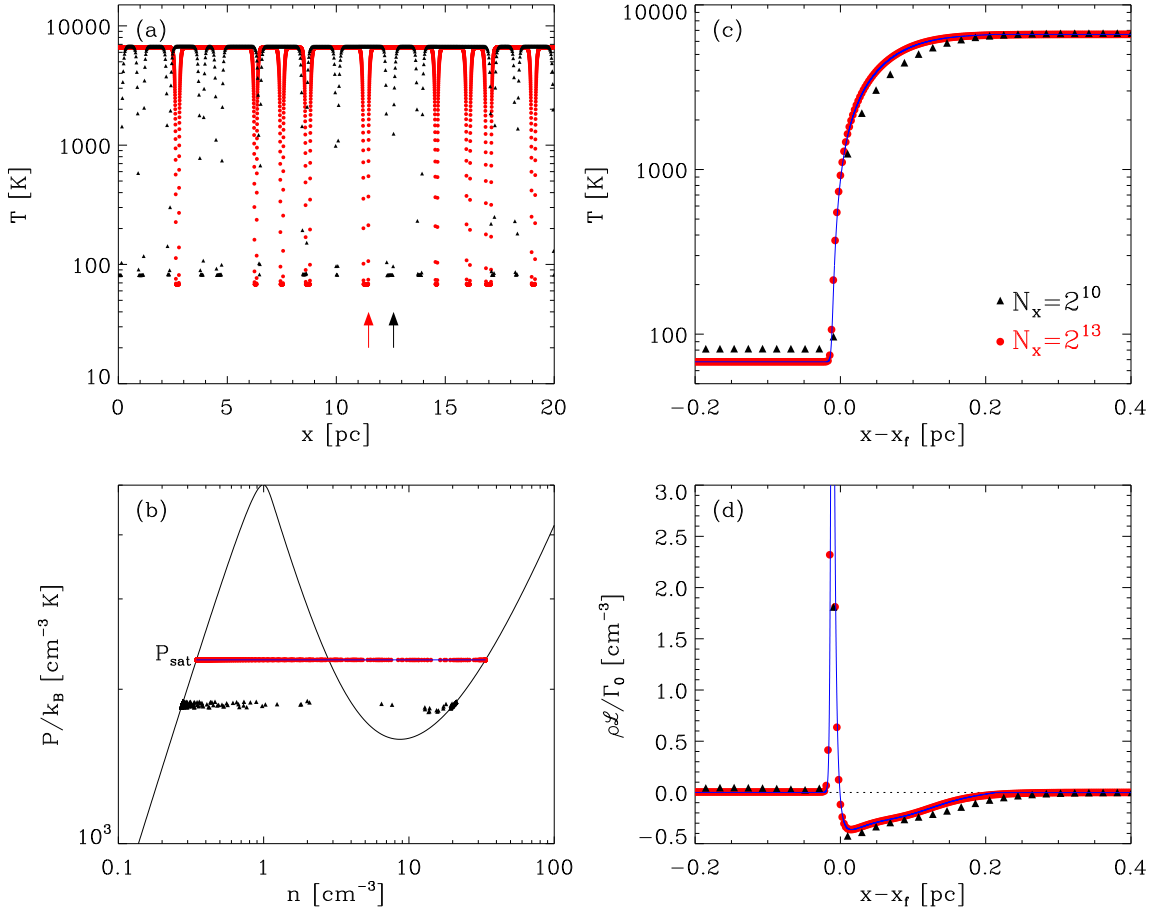}
  \caption{Comparisons between the low-resolution run with $N_x=2^{10}$ ($\Delta
    x=1.95\times10^{-2}\pc$; black triangles) and the
    high-resolution run with $N_x=2^{13}$ ($\Delta x=2.4\times10^{-3}\pc$;
    red circles) of the nonlinear static equilibrium at $t=500\Myr$
    obtained from TI for (a) overall temperature distributions, (b)
    scatter plots in the $n$--$P$ plane, with $\Psat$ equal to the
    analytic saturation pressure, and the profiles of (c) temperature and
    (d) heat-loss function near the CNM-WNM interface marked by the
    black or red arrow in (a). In (c) and (d), the solid lines draw
    the analytic predictions at $P=\Psat$.}\label{f:sat}
\end{figure*}

We, however, find that the density and temperature profiles, at the
saturated state of TI, of the interfaces between the CNM and WNM and
the corresponding equilibrium pressure are dependent upon numerical
resolution. To illustrate this, we compare in Figure \ref{f:sat} the
results for (a) temperature distribution and (b) scatter plots in
the $n$--$P$ plane at $t = 500 \Myr$ from the runs with $N_x =
2^{10}$ (with the grid spacing of $\Delta x=1.95\times10^{-2}\pc$;
black triangles) and $2^{13}$ (with $\Delta x=2.4\times10^{-3}\pc$;
red circles) zones.  Due to TI, the perturbations grow into a highly
nonlinear state where cold clumps are surrounded by a warm
intercloud gas. Some clumps merge together into larger ones at late
times, and the system reaches a quasi-steady state at around $50
\Myr$. Note that the high-resolution model recovers the saturation
pressure $\Psat=2282\kB\punit$ discussed in Section \ref{s:equil}
almost exactly with a root-mean-square velocity of $\delta v_{x}
\lesssim 0.01 \ms$. In the low-resolution model, on the other hand,
$P \approx 1870 \kB\punit$ with small fluctuations; the
corresponding velocity field has $\delta v_{x} = 14.0 \ms$, showing
that the velocity dispersion induced by TI also depends on numerical
resolution.  Figure \ref{f:sat}c,d directly compares the profiles of
(c) temperature and (d) heat-loss function across a CNM-WNM
interface between the low- and high-resolution models. The parts
indicated by the black and red arrows in Figure \ref{f:sat}a are
enlarged and shifted so as to make the front position $x_{\rm f}$
coincide. The solid lines representing the solution of Equations
\eqref{e:nCool} and \eqref{e:stn} at $P=\Psat$ are almost identical
to the results of the high-resolution run, while they deviate
considerably from those of the low-resolution model.  The
equilibrium pressure and velocity dispersion averaged over $t=
200$--$500 \Myr$ are listed in Columns (4) and (5) of Table
\ref{t:TI}.

The discrepancies of the saturation pressure and interface profiles in
our low-resolution runs from the analytic predictions are a numerical
artifact caused by overcooling in the cooling-dominated region. Figure
\ref{f:sat}d shows that strong radiative cooling is highly localized
to a narrow layer where temperature changes steeply. Its thickness is
$\sim 10^{-2}\pc$, comparable to the Field length
$(\FLone\FLtwo)^{1/2}$ in the thermally unstable medium. On the other
hand, the heating zone is relatively widely distributed over $\sim
0.1\pc$, comparable to $\FLtwo\sim \mu\FLone$. The $N_x=2^{10}$ model
with $\Delta x =0.02\pc$ resolves the heating zone quite well, but is
unable to resolve the cooling zone.  This results in net overcooling
across the interface, and thus reduction in the equilibrium pressure
(e.g., \citealt{pio04}). This in turn leads to larger temperatures and
lower densities of the CNM than the values at the saturation pressure
as seen in \citet{cho12}.

\begin{figure}
  \epsscale{1.2} \plotone{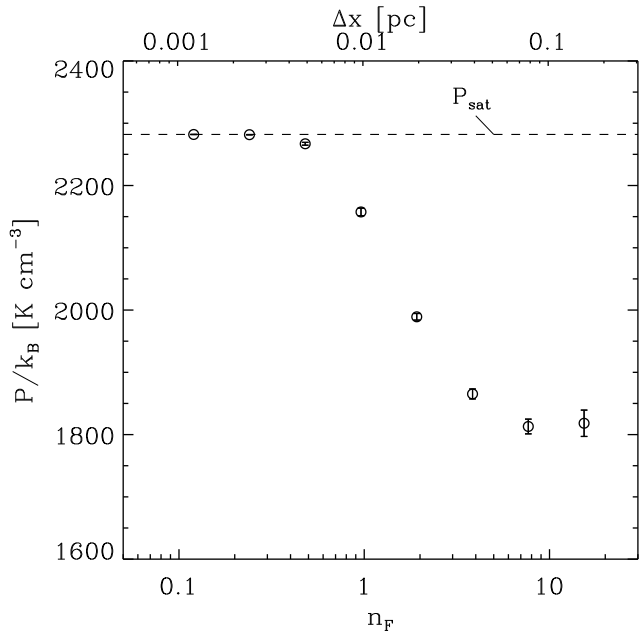}
  \caption{Dependence of numerically-found equilibrium pressure on the
    grid spacing (top $x$-axis) and on the Field number $n_{\rm F}
    \equiv \Delta x/ (\FLone\FLtwo)^{1/2}$ (bottom $x$-axis). Circles
    and errorbars indicate the mean values averaged over
    $t=200$--$500\Myr$ and the standard deviations. The
    dashed line marks $\Psat$ which is numerically attained only when
    $n_{\rm F} \simlt 0.25$.}\label{f:res}
\end{figure}

In Figure \ref{f:res} we plot the dependence on numerical resolution
of the equilibrium pressure obtained from our one-dimensional test
runs: circles and errorbars represent the mean values and standard
deviations over $t=200$--$500 \Myr$. Note that the grid spacing is
shown as the physical length in the top $x$-axis, while it is in terms
of the Field number $n_{\rm F} \equiv \Delta x/(\FLone\FLtwo)^{1/2}$
in the bottom $x$-axis. It is apparent that the equilibrium pressure
converges to $\Psat$ (the dashed line) as $n_{\rm F}$ decreases. Note
that $P\approx \Psat$ only when
\begin{equation}\label{e:res}
  n_{\rm F} \simlt 0.25\,,
\end{equation}
suggesting that it is necessary to resolve the Field length of the
transition layer by at least four zones in order to obtain accurate
solutions for CNM-WNM interfaces. In what follows, we present the
results of two-dimensional simulations for the DLI that satisfy the
condition \eqref{e:res}.

\subsection{Two-Dimensional Simulations}\label{s:2D}

We now turn to our central problem: the nonlinear evolution of the
DLI. We first restrict ourselves to the most unstable mode, and study
its linear and nonlinear growth as well as various physical properties
at saturation in detail. We then explore the case with multi-mode
perturbations.

\subsubsection{Single-Mode Case}\label{e:single}

As a background state, we select an evaporation front attained at
$\Peq/\kB = 1800 \punit$ as our fiducial model, and study its response
to the most unstable mode of the DLI.  The results of other
equilibrium state at different $\Peq$ are qualitatively similar.

The initial state is generated using the method described in Section
\ref{s:equil} for a given pressure, placing the evaporating front at
$x=0$. As our computation domain, we take a rectangular box that
spans $-(3/5)L_x\leq x\leq (2/5)L_x$ and $0\leq y\leq L_y$. The box
size is chosen as $L_x\times L_y = 5.64 \pc \times 1.06 \pc$, which
is large enough in the $x$-direction to encompass the asymptotic
regions of the flows and equals $\lmax$ in the $y$-direction. We set
up a $N_x\times N_y=2048 \times 384$ Cartesian grid with the cell
size of $\Delta x=\Delta y=2.8 \times 10^{-3} \pc$, which has the
Field number $n_{\rm F}= 0.20$, fulfilling the resolution
requirement of Equation \eqref{e:res}. For the boundary conditions,
we implement the inflow boundary condition at the left $x$-boundary
in which the density and velocity are set equal to the unperturbed
values every time step. This is not only to reduce the effects of
reflection of outgoing waves at the boundary but also to make the
upstream region at far field retain its unperturbed state.  We
impose the outflow boundary condition at the right $x$-boundary, and
the periodic conditions at the $y$-boundaries. The model parameters
and simulation outcomes of the fiducial model (Model MU69) are
listed in the top row of Table \ref{t:model}.

\begin{deluxetable*}{lccccccc}
\tabletypesize{\footnotesize} \tablewidth{0pt} \tablecaption{Model
Parameters and Simulation Outcomes
\label{t:model}} \tablehead{ %
\colhead{Model}           & %
\colhead{$\mu$}           & %
\colhead{$v_{x1}$}         & %
\colhead{$N_x\times N_y$}  & %
\colhead{$L_x\times L_y$}  & %
\colhead{$\Delta x$}       & %
\colhead{$D_s$}            & %
\colhead{$\mathcal{E}/(\jx L_y)$}     \\
\colhead{(1)}  & \colhead{(2)} & \colhead{(3)} & \colhead{(4)} &
\colhead{(5)}  & \colhead{(6)} & \colhead{(7)} & \colhead{(8)} }
\startdata
MU69   &  68.9 &  114  & 2048 $\times$ 384 & 5.64 $\times$ 1.06 & $2.8\times 10^{-3}$ & 0.54  & 2.44\\
MU69mul&  68.9 &  114  & 3072 $\times$ 1536& 8.46 $\times$ 4.23 & $2.8\times 10^{-3}$ & 0.59  & 2.49 \\
\hline
MU38   &  37.5 &  256  & 2048 $\times$ 384 & 6.33 $\times$ 1.18 & $3.5\times 10^{-3}$ & 0.45 & 2.00 \\
MU11   &  11.2 &  404  & 2048 $\times$ 384 & 10.5 $\times$ 1.97 & $9.6\times 10^{-3}$ & 0.43 & 1.44 \\
MU03   &  3.24 &  834  & 1024 $\times$ 384 & 9.22 $\times$ 3.46 & $2.6\times 10^{-2}$ & 0.33 & 1.10 %
\enddata
\tablecomments{Column 1: Model name. Columns 2--3: the expansion
factor and the CNM velocity at far upstream (cm s$^{-1}$) of the
background state. Columns 4--6: Number of zones, domain size (pc
$\times$ pc), and zone spacing (pc) of the simulation. Column 7--8:
the distortion amplitude (pc) and the evaporation rate relative to
the initial value at nonlinear saturation.}
\end{deluxetable*}

\begin{figure}
  \epsscale{1.1} \plotone{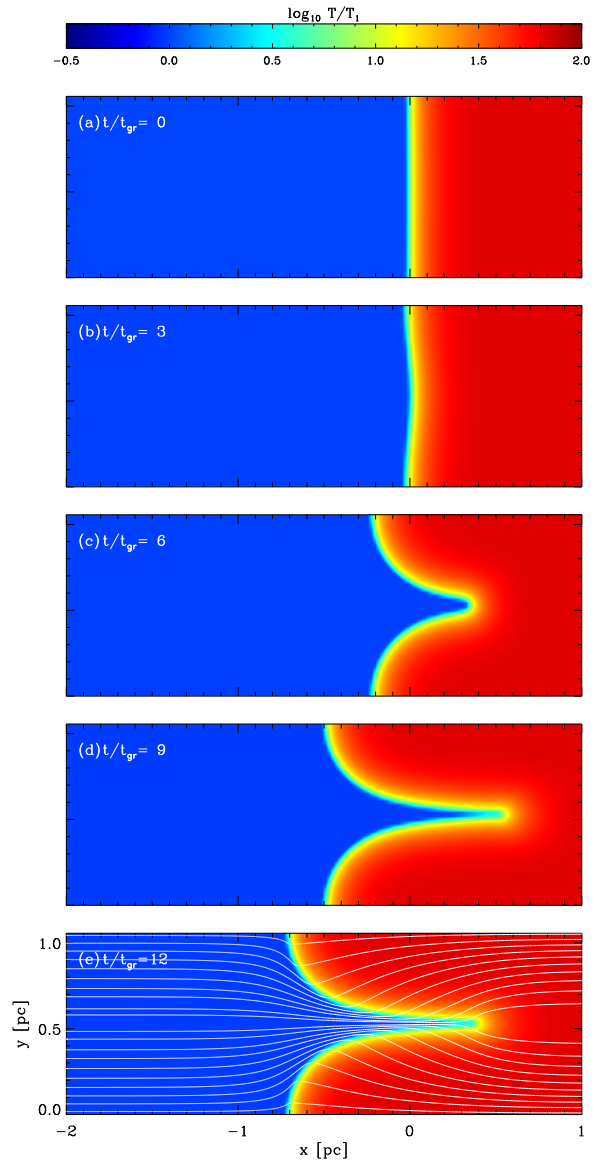}
  \caption{Snapshots of temperature distribution in logarithmic scale
    at $t/\tgrow=0$, 3, 6, 9, and 12 of Model MU69 that starts from
    single-mode perturbations with $L_y=\lmax$. The DLI grows
    exponentially at $t/\tgrow\simlt 6$ to distort the front, leading
    to a finger-like shape pointing downstream in the nonlinear
    stage.  There is no change in the front shape other than a
    translational shift between (d) and (e), suggesting a saturation
    of the DLI with an enhanced evaporation rate. In (e), gas
    streamlines are overlaid to show steady flow structures across the
    distorted front.}\label{f:contour}
\end{figure}

We have first checked that the initial front structure remains
stationary over a long period of time in the absence of any
perturbation. This confirms that our realization of the equilibrium
configuration is in a steady state. Next, we add small perturbations
to the initial configuration by shifting the front position slightly
as $\xf = -D_s(0) \cos(2\pi y/L_y)$ with the initial displacement
amplitude of $D_s(0)/L_y=10^{-3}$, which seeds the most unstable DLI
mode. We run the simulation until $700 \Myr$, corresponding to
$\sim17\tgrow$, well beyond the nonlinear saturation of the DLI.

\begin{figure}
  \epsscale{1.2} \plotone{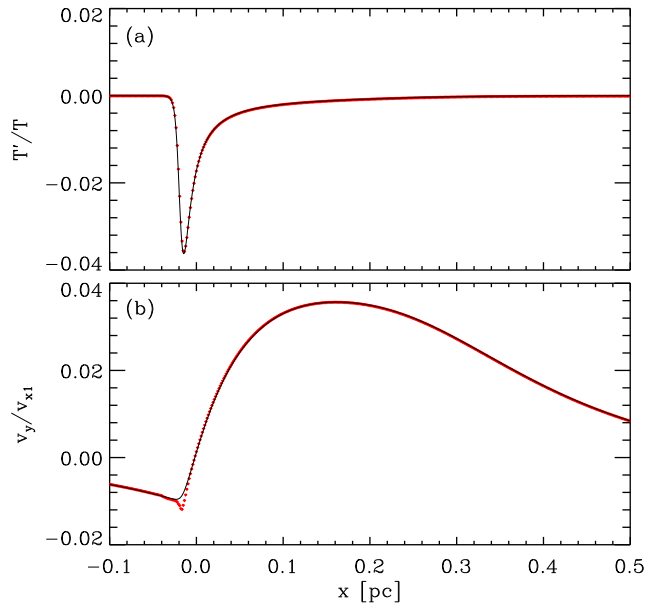}
  \caption{Distributions of (a) the perturbed temperature $T^\pr$
    along the $y=0.5L_y$ cut relative to the initial temperature $T$ and
    (b) perturbed velocity $v_y$ along the $y=0.25L_y$ cut relative to $v_{x1}$
    at $t/\tgrow=2$ of Model MU69 shown as
    dots. The results of the linear stability analysis are compared as
    solid lines.}\label{f:eigen}
\end{figure}

Figure \ref{f:contour} displays temperature snapshots in logarithmic
scales at $t/\tgrow=0$, 3, 6, 9, 12 of Model MU69. The characteristic
distortion of the front becomes noticeable at $t/\tgrow=1$, which is
growing exponentially with time. Figure \ref{f:eigen} compares the
numerical results (dots) for the perturbed temperature
$\Tpt(x,y=0.5L_y)$ and $y$-velocity $v_y (x,y=0.25L_y)$ at
$t/\tgrow=2$ with the analytic eigenfunctions (solid lines) obtained
from the linear stability analysis. At this time, the system is still
in the linear regime. The agreement between the numerical results and
the predictions of the linear theory is excellent. We note that the
profile of the perturbed temperature closely resembles that of the
initial temperature gradient, i.e., $\Tpt \propto -dT/dx$ in the
linear regime., as proven by \citet{lib94}.

To describe the front shape at arbitrary $t$, we introduce a curve
$\mathcal{C}_{\rm{f}} = x - s(y,t)=0$, where $s(y,t)$ denotes the
$x$-position of the isotherm with $T_{\rm f}=\sqrt{T_1T_2}$ at given
$x$ and $t$. Then, the evaporation rate of the CNM per unit area is
given by
\begin{equation}
 \epsilon = \rho_{\rm f} \left(
 \mathbf{\hat{s}}\cdot \mathbf{v}_{\rm{f}}
 -\frac{1}{|\nabla \mathcal{C_{\rm f}}|}\frac{\partial s}{\partial t} \right),
\end{equation}
where $\mathbf{\hat{s}} = \nabla\mathcal{C}_{\rm f} / |
\nabla\mathcal{C}_{\rm f}|$ is the unit vector normal to the front
directed towards the WNM, $\rho_{\rm{f}}$ and $\mathbf{v}_{\rm{f}}$
denote the gas density and velocity at the front, respectively.  The
total evaporation rate in the computational domain is then
$\mathcal{E} =\int_{\mathcal{C}_{\rm f}} \epsilon dl$, where the
integration is carried along the front.  Note that $s=\text{constant}$
for a vertically-straight, steady front, yielding $\epsilon_0 = \jx =
2066\; \overline{m}\cm^{-2} \;\rm{s}^{-1}$ and $\mathcal{E}_0=\jx L_y$
in the unperturbed state.

\begin{figure*}
  \epsscale{1.}
  \plotone{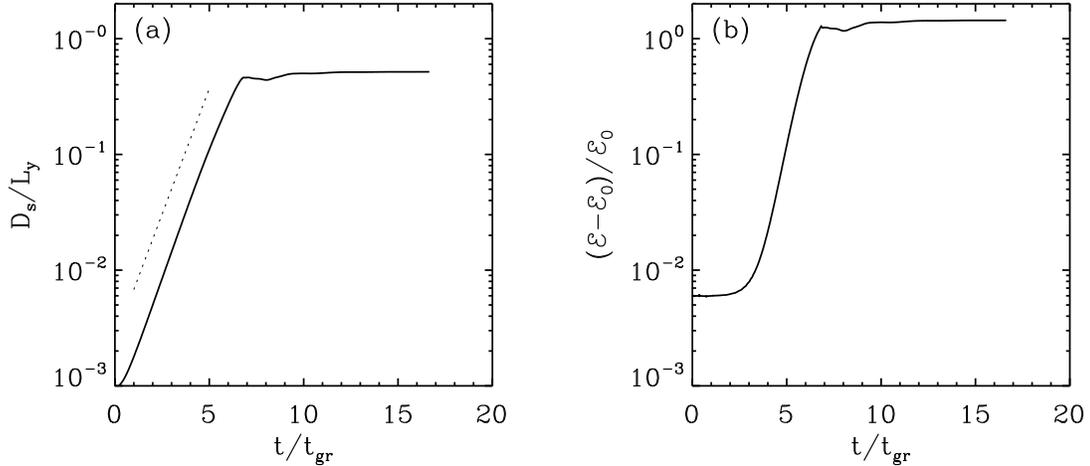}
  \caption{Temporal changes of (a) the distortion amplitude $D_s$ of
    the front and (b) the total evaporation rate $\mathcal{E}$ in
    Model MU69. The predicted linear growth rate is indicated as a
    dotted line segment in (a). The DLI saturates at $t/\tgrow\sim
    6$--$7$, with the evaporation rate enhanced by a factor of
    $2.4$.}\label{f:mucomp}
\end{figure*}

Figure \ref{f:mucomp} plots the temporal changes of (a) the distortion
amplitude $D_s \equiv (\max(s) - \min(s))/2$ and (b) the total
evaporation rate in Model MU69. The DLI grows exponentially at early
time, whose rate is consistent with the linear-theory prediction
plotted as a short dotted line. As Figure \ref{f:contour} shows, the
front becomes increasingly more distorted as the DLI grows, and has
$D_s$ comparable to $\lmax$ at $t/\tgrow\sim6$. The front eventually
develops into a finger-like structure that protrudes toward the
downstream direction.  The distorted front does not grow further after
$t/\tgrow \sim7$, indicating that the DLI saturates nonlinearly. The
front distortion increases the length of the front where the inflowing
CNM turns into the WNM. Thus, the growth of the DLI inevitably results
in an increase in the evaporation rate, which in turn causes the
distorted front to move toward the upstream direction in our
simulation, as evidenced in Figure \ref{f:contour}.

According to \citet{zel66}, the nonlinear saturation of the flame
instability occurs due to the Huygens principle which states that
every point of the front can be regarded as a source of a secondary
spherical wave (see also \citealt{zel85}). Suppose a curved front that
is moving relative to the CNM. Since waves launched from the concave
(convex) parts of the front to the WNM diverge (converge), the convex
parts eventually develop cusps in the limit of infinitesimally-thin
front. The propagation velocity of the convex parts is larger than
that of the concave part, which balances the growing tendency of the
distortion amplitude of the front, resulting in a steady
configuration. Smoothed by thermal conduction, the distorted front in
our model does not display a sharp cusp.

Figure \ref{f:contour}e plots gas streamlines (white lines) around
the front at $t/\tgrow = 12$ in Model MU69.  Although the refracted
flow field indicates a production of some vorticity at the distorted
front ($|\nabla\times \mathbf{v}| \simlt 1.79 \times
10^{-14}\;\rm{s}^{-1}$), the DLI of an initially laminar flow does
not lead to turbulence at nonlinear saturation, consistent with the
results of \citet{bel04a}. The local evaporation rate $\epsilon$
varies along the front in such a way that it is largest ($\sim
4.68\jx$) at the tip of the finger due to the largest curvature and
hence the efficient conductive heating from the surrounding WNM, and
becomes smallest ($\sim 0.77\jx$) at the wing sides. When integrated
over the front length, the total evaporation is $\mathcal{E}= 2.4
\jx L_y$ at saturation, 2.4 times larger than $\mathcal{E}_0$. This
increase of $\mathcal{E}$ is in complete accordance with the larger
inflow velocity of the CNM relative to the front, which is $\sim
2.4$ times larger than the initial plane-parallel value. Because the
fractional increase in the front length is also a factor of $2.4$ in
the saturated state, the increase in the total evaporation rate is
due directly to the increase in the front length (see also Section
\ref{s:mu}).

\subsubsection{Multi-mode Perturbation}

\begin{figure*}
  \plotone{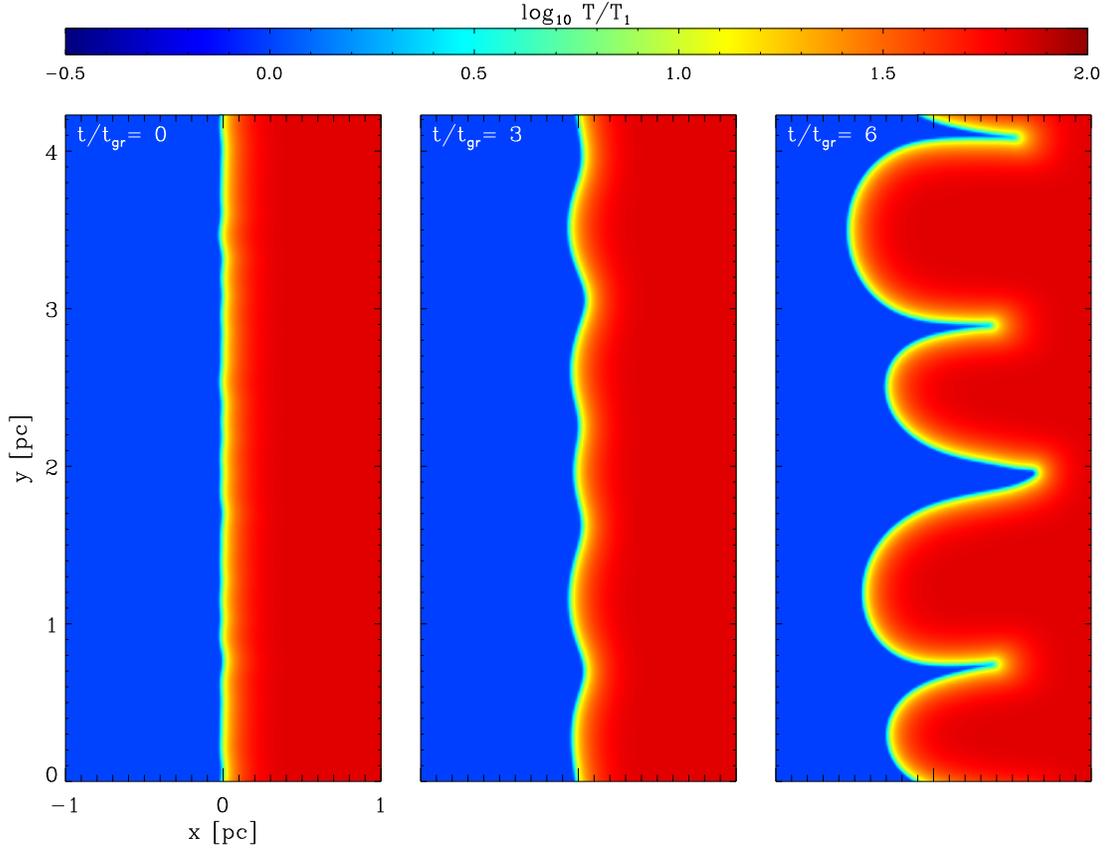}
  \caption{Temperature snapshots at $t/\tgrow=0$, 3, and 6 in Model
    MU69mul with $L_y=4\lmax$ that starts from random
    perturbations. While various modes grow in the linear stage, the
    system at late time is dominated by the most unstable mode that
    produces four finger-like structures.}\label{f:multi}
\end{figure*}

We also run Model MU69mul that has a larger simulation domain with
$L_x \times L_y =$ $8.46 \pc$ $\times$ $4.23 \pc$ than in the
single-mode case, which can accommodate perturbations with wavelength
up to $4\lmax$ (see the second row of Table \ref{t:model}). This model
is to explore whether the system readily picks up the most unstable
mode of the DLI.  We take the same background state as in Model MU69,
and displace the front positions in $x$ randomly with amplitude of
$10^{-3}L_y$ from the equilibrium location. Figure \ref{f:multi}
displays temperature snapshots in logarithmic scale at $t/\tgrow=0$,
3, and 6. Perturbations grow at rates depending on their
wavelengths. At early time, perturbations with large initial
amplitudes emerge first, which happen to be the mode with
$m=L_y/\lambda=5$ at $t/\tgrow=3$. But, it is eventually the most
unstable $m=4$ mode that dominates to form finger-like nonlinear
structures at late time. Although the growth of other unstable modes
makes the spacing between the fingers irregular to some extent, the
overall morphology and the increase in the evaporation rate at
nonlinear saturation are consistent with the results of Model MU69. We
again note that the ratio of the kinetic energy to the thermal energy
in Model MU69mul at $t/\tgrow=6$ is $4 \times 10^{-4}$ and the system
remains laminar, without evolving into a turbulent state.

\subsubsection{Effects of Density Ratio}\label{s:mu}

As shown in the preceding sections, the DLI of an evaporation front in
the ISM leads to front deformation and an increase in the evaporation
rate, without driving turbulence. This is qualitatively consistent
with the results of numerical studies of the DLI in the context of C/O
thermonuclear flames in Type Ia supernovae (e.g., \citealt{bel04a})
and combustion in heat engines (e.g., \citealt{byc96,tra99}). However,
our results differ quantitatively in that while the front deformation
and the associated increase of the flame propagation speed are only a
few percents in the other studies, they are more than $100\%$ in our
simulations.  These differences are most likely caused by differences
in the expansion factor $\mu$ between the models. Note that $\mu\simlt
10$ in terrestrial flames and $\mu\simlt 2$ in SN thermonuclear
flames, which is about an order of magnitude smaller than that of
evaporation fronts in the ISM.

\begin{figure*}
  \plotone{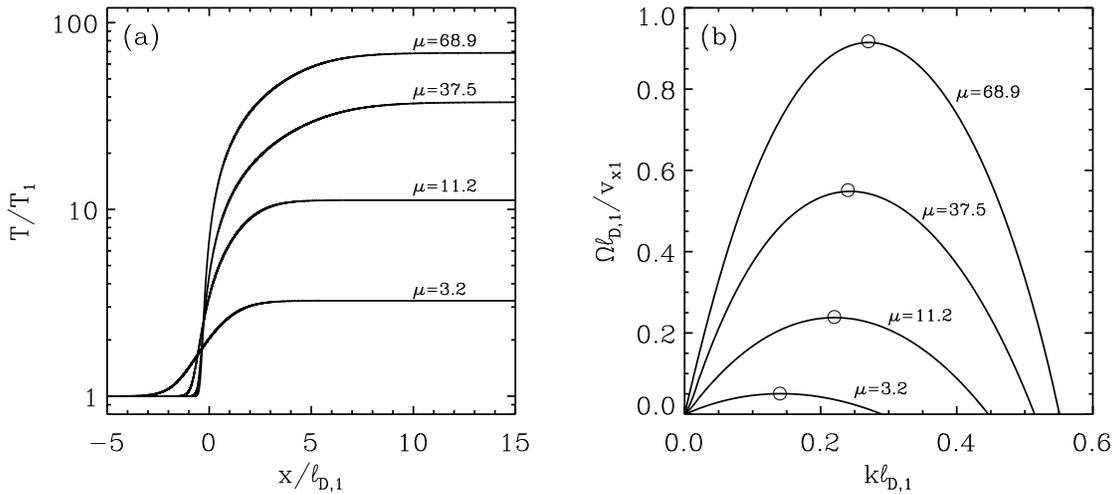}
  \caption{(a) Temperature profiles of evaporation fronts and (b)
    dispersion relation of the DLI for the modified heating rate with
    $n_0=\infty$, 10, 4, and $2\pcc$, or the corresponding expansion
    factor of $\mu=68.9$, 37.5, 11.2, and 3.2, from top to bottom. In
    (b), circles indicate the growth rates measured from numerical
    simulations in the linear phase, in good agreement with the
    linear-theory results.}\label{f:mugr}
\end{figure*}

\begin{figure}
  \epsscale{1.2}
  \plotone{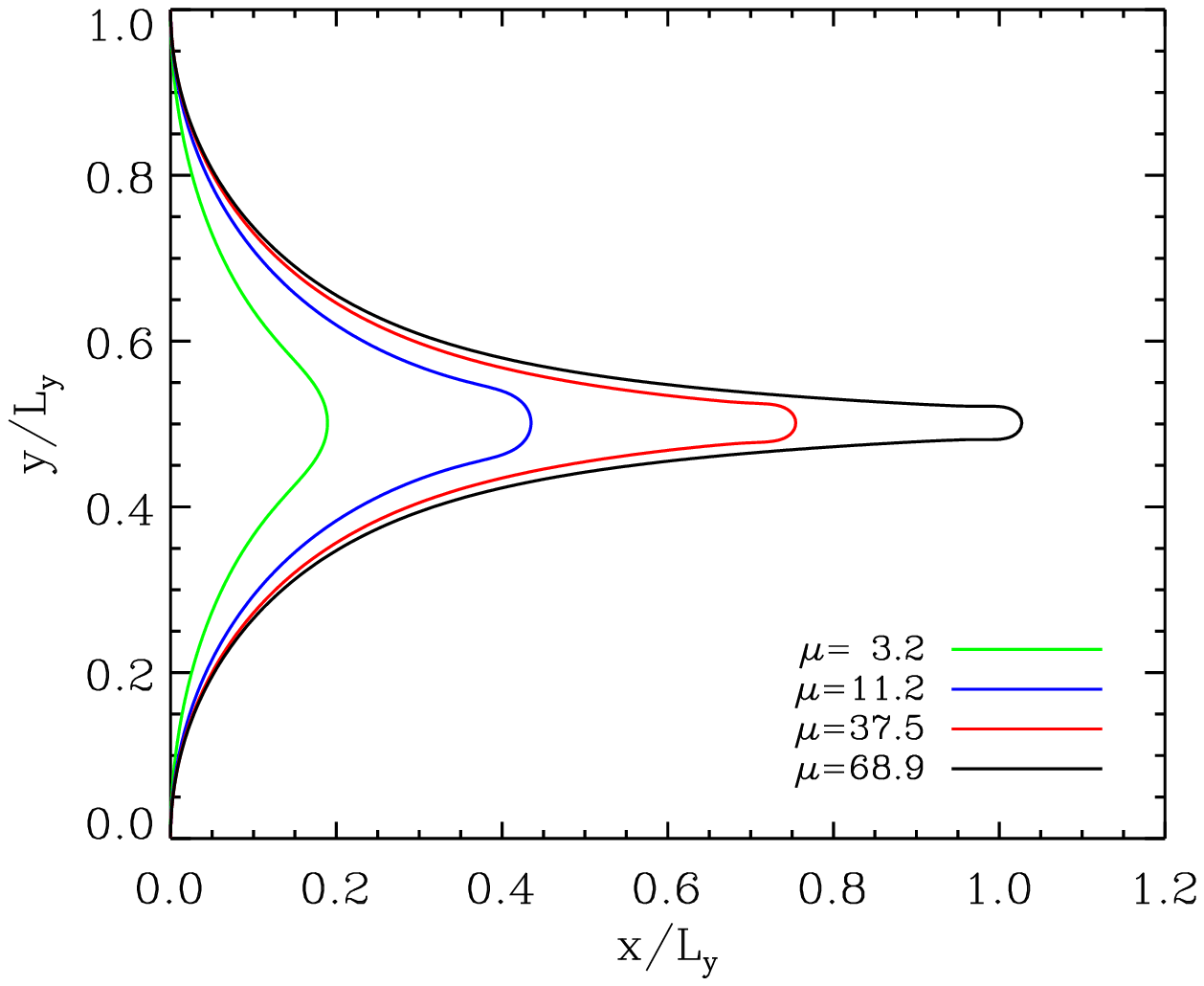}
  \caption{Comparison of the front shapes at $t/\tgrow=12$ when the
    DLI achieves nonlinear saturation in models with differing $\mu$
    under the modified heating function.  A model with smaller $\mu$
    has a smaller distortion amplitude and a lower evaporation rate
    at saturation.}\label{f:muf}
\end{figure}

To directly assess the impact of varying $\mu$ on the nonlinear
saturation of DLI, we conduct simulations of heuristically modeled
fronts with smaller $\mu$. For this purpose, we modify the
density-independent heating rate to
\begin{equation}
 \Gamma = \Gamma_0 \times \dfrac{\exp[(n/n_0)^3]}{1+n/n_0}\,,
\end{equation}
where $n_0$ is a free parameter. Note that $\Gamma\rightarrow
\Gamma_0$ as $n_0\rightarrow\infty$. For finite $n_0$, $\Gamma$
increases rapidly with $n\simgt n_0$, lowering the  equilibrium CNM
density without much effect on the WNM density. Thus, a smaller
value of $n_0$ results in smaller $\mu$ in an equilibrium
configuration.

Figure \ref{f:mugr}a plots a few equilibrium density profiles for
$n_0=10$, 4, and $2\pcc$; the corresponding density contrasts,
pressures, and the far upstream inflow speeds are $\mu=38$, 11, 3.2,
$\Peq/\kB=1500$, 2000, $2800\punit$, and $v_{x1}=256$, 404,
$834\cms$, respectively. The case with $n_0=\infty$ is also plotted
for comparison. For these steady fronts, we perform the linear
stability analysis and plot the resulting growth rates in Figure
\ref{f:mugr}b. Clearly, the DLI grows slower with decreasing $\mu$,
which is expected from Equation \eqref{e:Omega0}. The critical
wavenumber becomes smaller with decreasing $\mu$.

For these evaporation fronts with the modified heating rate, we run
numerical simulations of the DLI by taking $L_y$ equal to the
wavelength of the fastest growing mode. The parameters for these runs,
named MU38, MU11, and MU03, are given in Table \ref{t:model}.  Note
that the models with small $\mu$ easily meet the resolution
requirement because of the higher CNM temperature (leading to the
increase of $\FLone$). In all models, we displace the fronts
sinusoidally with amplitude of $10^{-3}L_y$. As in Model MU69, the
evaporation fronts in these lower-$\mu$ models are increasingly more
distorted with time as a result of the DLI. In Figure \ref{f:mugr}b,
we mark as open circles the growth rates measured from the simulations
in the linear stage, in good agreement with the results of the linear
stability analysis. The DLI soon enters the nonlinear regime and
ceases to grow further after $t/\tgrow \sim 7$, although models with
larger $\mu$ tend to saturate later because of a larger amplification
factor.

Figure \ref{f:muf} compares the front shapes from models with
different $\mu$ at $t/\tgrow = 12$. Apparently, the distortion
amplitude and the evaporation rate at saturation become smaller with
decreasing $\mu$. For example, the ratio of the distortion amplitude
to the wavelength of the most unstable mode is $\sim 0.095$ and $0.38$
in Models MU03 and MU38, respectively. The total evaporation rate at
saturation can be fitted by
\begin{equation}\label{e:evap}
 \mathcal{E}/\mathcal{E}_0 \approx 1 + 0.41 (\log \mu)^2\,,
\end{equation}
for $3\leq \mu\leq 70$. Extrapolating this result to $\mu=1.52$
corresponding to C/O thermonuclear flames, we obtain
$(\mathcal{E}-\mathcal{E}_0)/\mathcal{E}_0 =1.4\%$, roughly consistent
with the result of \citet{bel04a}. Again, the evaporation fronts in
all models do not develop a cusp at the location of the maximum
distortion due to the smoothing effect of thermal conduction.

\section{Summary and Discussion}\label{s:sum}

While the behavior of a thermally-bistable fluid consisting of the CNM
and WNM is becoming increasingly more important for numerical studies
of the ISM, relatively little attention has been directed to dynamics
of their interfaces. In this paper we have presented the results of
the full linear stability analysis and numerical simulations for the
corrugational instability, or the DLI, of evaporation fronts in the
ISM.  As an unperturbed state, we consider an evaporation front in
plane-parallel geometry and take a constant value for thermal
conductivity. Our key findings are summarized as follows.

1. The type and structure of a thermal front between the CNM and WNM
in steady equilibrium are determined by the equilibrium thermal
pressure $\Peq$, such that the front becomes a condensation front when
$\Psat<\Peq<\Pmax$ across which the WNM changes to the CNM, and an
evaporating front when $\Pmin<\Peq<\Psat$, where $\Psat$ is the
saturation pressure for a static front and $\Pmax$ and $\Pmin$ refer
to the maximum and minimum pressures for two-phase equilibrium,
respectively. For our adopted heat-loss function and thermal
conductivity, $\Psat/\kB=2282\punit$, $\Pmax/\kB=5005\punit$, and
$\Pmin/\kB=1597\punit$. The incident velocity $v_{x1}$ of the CNM at
far upstream relative to the evaporating front is limited to below
$250\cms$, much smaller than the sound speed, making the isobaric
approximation valid. The overall front width is comparable to the
Field length in the WNM.

2. We perform the full linear stability analysis of the DLI in the
presence of thermal conduction following the eigenvalue approach of
\citet{lib94}. While the front thickness is determined by the Field
length, the length and time scales of the instability are well
characterized by the diffusion length $\DL$ and the corresponding
crossing time $\DL/v_{x1}$ (see Fig.\ \ref{f:gr} and Eqs.\
\eqref{e:maxlen} and \eqref{e:maxtime}). The linear dispersion
relations show that perturbations with $\lambda/\DLone \simlt 12$ are
completely stabilized by conduction, while they are well approximated
by Equation \eqref{e:Omega0} for very long-wavelength perturbations.
The growth rate also depends on the expansion factor $\mu$ defined by
the density ratio of the CNM to WNM.

3. Using one-dimensional simulations of TI, we demonstrate that it is
important to resolve the Field length of a transition layer between
the CNM and WNM by at least four grid points in order to obtain
accurate density and temperature distributions as well as the correct
saturation pressure (see Eq.\ \eqref{e:res}). Otherwise, the region of
strong radiative cooling near the CNM would be unresolved, giving rise
to overcooling and reduction in the equilibrium pressure.

4. Two-dimensional simulations of the DLI of an evaporating front show
that small perturbations grow exponentially in the linear regime to
bend the front, and saturate nonlinearly typically at
$t/\tgrow\sim7$. The numerical growth rates in the linear stage are in
good agreement with the predictions of the linear theory. In the
nonlinear regime, the front is in a steady state and has a finger-like
shape pointing toward the WNM, without developing turbulent flows. The
presence of thermal conduction smooths out the front that would
otherwise be cuspy with infinitesimal front thickness. The increase in
the front length at saturation directly translates into an increase in
the evaporation rate. For our fiducial model with
$\Peq/\kB=1800\punit$ and $\mu=68.9$, the saturated evaporation rate
is increased by a factor of $\sim2.4$ relative to the initial
plane-parallel value. By running control models with the modified
heating rate, we find that the evaporation rate at saturation relative
to the initial value depends on $\mu$ and is given by Equation
\eqref{e:evap}.

The importance of resolving interfaces accurately has been
emphasized by a number of authors in various contexts. For example,
\citet{koy04} ran various simulations of TI with differing
resolution, and found that numerical convergence for the density
distribution is achieved only when the grid size is less than one
third of the local Field length, similarly to our results.
\citet{kru07} discussed numerically-induced cooling in an ionization
front advancing into a surrounding molecular medium. By comparing
their numerical results with the analytic solutions, they showed
that overcooling arises if the size of a computational cell is
larger than the true thickness of the front, slowing down the front
expansion. Overcooling is due to numerical mixing, leading to the
overestimation of the amount of molecular gas around the front, a
far more efficient coolant than ions and atoms. An analogous
situation takes place in cosmological simulations that often lack
sufficient resolution to resolve interfaces between gases of
different temperatures, giving rise to the classical overcooling
problem \citep[e.g.,][]{kaz92,mac13}. In our models, unresolved
cooling in the interfaces leads to non-vanishing $Q$ even for a
static front, which results in non-zero gas motions whose speed is
roughly $\sim \FLen /t_{\rm{cool}}$ from Equation \eqref{e:jQ} (see
also \citealt{iwa12}).

As represented by Equations \eqref{e:maxlen} and \eqref{e:maxtime},
the time and length scales of the DLI depend on the inflow speed
$v_{x1}$ and the density $n_1$ of the CNM, which in turn depends
rather sensitively on the adopted heat-loss function. In this work, we
considered the ISM parameters representing the solar neighborhood
conditions and found that the most unstable mode has a typical
wavelength of $\sim 1\pc$ and a growth time of $\sim50\Myr$ for $n_1 =
10 \pcc$, $v_{x1} = 1 \ms$, and $\kappa = 10^5 \condunit$. However,
the radiative cooling and heating may vary considerably in space and
time, depending on local conditions such as star-forming activity, gas
column density, abundances of the main coolants, etc., all of which
can affect the density and temperature profiles of an evaporating
front in equilibrium.  In the inner regions of a galactic disk, for
example, elevated star formation rates together with stronger ionizing
background radiation lead to a significantly enhanced heating rate,
which would make the thermal equilibrium curve in the $n$--$P$ plane
shifted upward and rightward \citep[e.g.,][]{par03,wol03, cox05}. When
the heating rate is five times larger than the one we adopt in the
present work, for instance, an evaporation front has $v_{x1} = 254
\cms$ and $n_1 = 91 \pcc$ at $\Peq/\kB=9000\punit$.  The growth time
and wavelength of the fastest growing mode is then $\tgrow = 1.7 \Myr$
and $\lmax = 0.094 \pc$, suggesting that the growth of the DLI is
highly subject to environmental conditions.

\citet{ino06} examined the linear stability of evaporation fronts in
the ISM accounting for the effect of temperature-dependent
conductivity. They obtained the growth rate of the DLI by considering
only thermal modes, while neglecting incompressible and vortex modes
(see Appendix \ref{a:lin}).  The cooling function they adopted is
different from ours in that they took the first terms (due to
Ly$\alpha$ emissions) and second terms (due to \ion{C}{2} lines) in
Equation \eqref{e:heatfn3} about 30 times larger and smaller than
those in our paper, respectively. The resulting pressure range for
two-phase equilibrium is $637 \punit < P/\kB < 12600 \punit$, much
wider than ours. The typical inflow speed and density of the CNM in
their models are $v_{x1}=5\ms$ and $n_1=30\pcc$, larger by about a
factor of $5$ and $2$ than our values. They found that the most
unstable mode has a growth time of $\sim0.3 \Myr$ and a wavelength of
$\sim 0.1 \pc$. Although it is difficult to make a direct comparison
due to the differences in the cooling function and thermal
conductivity, our results are overall consistent with their results if
$v_{x1}$ and $n_1$ are taken appropriately in Equations
\eqref{e:maxlen} and \eqref{e:maxtime}.

In this work, we have investigated the DLI under initially laminar
conditions, showing that the DLI itself does not lead to turbulence in
the ISM, consistent with the results of the DLI in terrestrial flames
(e.g., \citealt{byc96,tra99}) and thermonuclear flames (e.g.
\citealt{rop03,bel04a,bel04b}). The real ISM, however, is shaped by
turbulence on a wide range of length scales.  Since the DLI involves
deflection of gas streamlines at the front, the presence of
non-uniform distribution of pressure and velocity in the background
flows may obstruct the development of the instability. \citet{hey04}
reported that the velocity dispersions of clouds behave as $v(\ell) =
v(\ell_0)(\ell/\ell_0)^q$, with $v(\ell_0=1\pc) = 0.9\kms$ and $q =
0.56$, from $30\pc$ down to $0.03\pc$ scales (see also
\citealt{mo07}). Assuming that the evaporating flow decouples from
turbulence, the DLI grows only if its growth time is larger than the
eddy turnover time at $\ell=\lmax$, or if $v_{x1} \simgt 15 \ms
(n_1/10\pcc)^{-q/(1+q)}(\kappa/\kappa_0)^{q/(1+q)}$ from Equations
\eqref{e:maxlen} and \eqref{e:maxtime}. For our choice of the
heat-loss function, $v_{x1}$ is less than $\sim 3 \ms$ for steady
evaporation fronts, suggesting that the DLI of CNM-WNM evaporation
fronts in the neutral ISM is unlikely to grow into the nonlinear
regime unless the interfaces are strongly protected from the ISM
turbulence.

While we in this work focused on evaporation fronts between the CNM
and WNM in a plane-parallel geometry, we briefly comment on the
evaporation of a spherical cloud bathed in a hot ISM. According to
\citet{mck77}, the evaporation rate of a cold cloud with radius $R = 1
\pc$ and density $n\sim10\pcc$ is estimated to be $\dot{M} = 1.5
\times 10^{26} \;\rm{g}\;\rm{yr}^{-1}$. The corresponding CNM velocity
with respect to the front is $v_{x1}\sim \dot{M}/(4\pi R^2
\overline{m}n) = 18 \ms$, indicating that the DLI is likely to grow
even in the presence of the ISM turbulence. In addition, the growth
time of the DLI based on our results is $0.35 \Myr$, much smaller than
the expected evaporation time scale $\sim 18 \Myr$, suggesting that
the evaporation front of a spherical cloud in a hot medium may suffer
from the DLI. Of course, the real assessment of the DLI in this
situation requires consideration of the curvature effect as well as
realistic thermal conductivity and heating/cooling rates applicable
for the hot phase that can substantially alter the background states
and evaporation processes.

\acknowledgments We are grateful to the referee for a helpful
report. This work was supported by the National Research Foundation
of Korea (NRF) grant funded by the Korean government (MEST), No.\
2010-0000712. The computation of this work was supported by the
Supercomputing Center/Korea Institute of Science and Technology
Information with supercomputing resources including technical
support (KSC-2012-C3-19).

\appendix

\section{Linear Stability Analysis}\label{a:lin}

We here present the method to obtain linear dispersion relations of
the DLI of evaporating fronts in the presence of thermal conduction.
Our approach essentially follows \citet{lib94} who studied instability
of terrestrial flames (see also \citealt{lib08}).

\subsection{Perturbation Equations}

We initially consider a one-dimensional steady evaporation front
located at $x=0$, like the one shown in Figure \ref{f:t0}, in which
the density, velocity, and temperature vary with $x$. We apply
two-dimensional perturbations to the steady configuration, and seek
exponentially growing modes. Assuming that the perturbation amplitudes
are small, Equations \eqref{e:cont}--\eqref{e:ideal} are linearized to
\begin{equation}\label{e:per1}
  \dfrac{\p \rho^{\pr}}{\p t} + \dfrac{\p}{\p x}(\rho v_x^{\pr} + \rho^{\pr}v_x) +
  \dfrac{\p}{\p y}(\rho v_y^{\pr}) = 0\,,
\end{equation}
\begin{equation}\label{e:per2}
  \dfrac{\p}{\p t}(\rho v_x^{\pr} + \rho^{\pr}v_x) +
  \dfrac{\p}{\p x}(P^{\pr} + 2\rho v_xv_x^{\pr} + \rho^{\pr}v_x^2) +
  \dfrac{\p}{\p y}(\rho v_x v_y^{\pr})
  = 0\,,
\end{equation}
\begin{equation}\label{e:per3}
  \dfrac{\p}{\p t}(\rho v_y^{\pr}) +
  \dfrac{\p}{\p x}(\rho v_xv_y^{\pr}) +
  \dfrac{\p P^{\pr}}{\p y} = 0\,,
\end{equation}
\begin{equation}\label{e:per4}
  \kappa \left(\dfrac{\p^2}{\p
      x^2} + \dfrac{\p^2}{\p y^2}\right) T^{\pr} =
  \rho c_P\dfrac{\p T^{\pr}}{\p t} + (\rho^{\pr}v_x + \rho v_x^{\pr}) c_P
  \dfrac{dT}{dx} + \rho v_x c_P\dfrac{dT^{\pr}}{dx} + \dfrac{d (\rho \mathcal{L})}{dT}
  T^{\pr}\,,
\end{equation}
and
\begin{equation}\label{e:per5}
  \frac{P^\pr}{P} = \frac{\rho^\pr}{\rho} + \frac{T^\pr}{T}\,,
\end{equation}
where the primes indicate the perturbed quantities. In deriving
Equation \eqref{e:per4}, we have made the isobaric approximation under
which $\rho\mathcal{L}$ is a univariate function of $T$.  The isobaric
assumption is valid since the fractional change of thermal pressure is
proportional to the square of the Mach number that is much less than
unity even for fastest evaporating flows (e.g., \citealt{lib08}).

It is convenient to take $j_x^{\pr} = \rho v_x^{\pr} + \rho^{\pr} v_x$
and $M^{\pr}_x = P^{\pr} + 2\rho v_xv_x^{\pr} + \rho^{\pr}v_x^2$,
instead of $\rho^{\pr}$ and $v_x^\pr$, as independent perturbed
variables. We decompose the perturbations as $\propto \exp(\Omega t +
iky)$, where $k$ and $\Omega$ denote the wavenumber and growth rate,
respectively. We introduce the dimensionless perturbed variables as
\begin{equation}
\left(
\begin{array}{c}
j_x^{\pr}/\jx \\
M^{\pr}/(\rho_1v_{x1}^2) \\
v_y^{\pr}/v_{x1} \\
T^{\pr}/T_1
\end{array}
\right)
=
\mathrm{Re}
\left[
\left(
\begin{array}{c}
\mathcal{J}^{\pr}(\xi) \\
\mathcal{M}^{\pr}(\xi) \\
-i\mathcal{V}^{\pr}(\xi) \\
\mathcal{T}^{\pr}(\xi)
\end{array}
\right) e^{\Omega t + iky} \right]\,,
\end{equation}
where the quantities with subscript ``1'' are evaluated at
$x=-\infty$. Then, Equations \eqref{e:per1}--\eqref{e:per5} can be
written as
\begin{equation}\label{e:nodim1}
\dfrac{d\mathcal{J}^{\pr}}{d\xi} = - \nu \dfrac{\mathcal{V^{\pr}}}{\mathcal{T}} +
\nu\sigma\dfrac{\mathcal{T}^{\pr}}{\mathcal{T}^2}\,,
\end{equation}
\begin{equation}\label{e:nodim2}
  \dfrac{d\mathcal{M}^{\pr}}{d\xi} =
  -\nu\sigma\mathcal{J}^{\pr} - \nu \mathcal{V}^{\pr}\,,
\end{equation}
\begin{equation}\label{e:nodim3}
  \dfrac{d\mathcal{V}^{\pr}}{d\xi} = - 2\nu \mathcal{T} \mathcal{J}^{\pr}  +
  \nu \mathcal{M}^{\pr} -
  \nu\sigma\dfrac{\mathcal{V}^{\pr}}{\mathcal{T}} - \nu \mathcal{T}^{\pr}\,,
\end{equation}
\begin{equation}\label{e:nodim4}
  \dfrac{d^2\mathcal{T}^{\pr}}{d\xi^2} - \dfrac{d\mathcal{T}^{\pr}}{d\xi} +
  \dfrac{d \mathcal{H}}{d\mathcal{T}}\mathcal{T}^{\pr} =
  \dfrac{d\mathcal{T}}{d\xi}\mathcal{J}^{\pr} + \nu\sigma\dfrac{\mathcal{T}^{\pr}}{\mathcal{T}} +
  \nu^2\mathcal{T}^{\pr}\,,
\end{equation}
where $\xi \equiv x/\DLone$, $\mathcal{T} \equiv T/T_1$, $\sigma
\equiv \Omega/(kv_{x1})$, $\nu \equiv k\DLone$, and
$\mathcal{H}\equiv -(\DLone/\FLone)^2\rho\mathcal{L}/(n_1\Gamma_0)$.
Let $\mathbf{U}(\xi)$ and $\mathbf{D}(\xi)$ denote the vectors,
$(\mathcal{J}^{\pr},\,\mathcal{V}^{\pr},\,
\mathcal{M}^{\pr},\,\mathcal{T}^{\pr},\,d\mathcal{T}^{\pr}/d\xi)$,
that describe the perturbations in the upstream and downstream sides
of the front, respectively. Our strategy is to first obtain
$\mathbf{U}$ and $\mathbf{D}$ by integrating Equations
\eqref{e:nodim1}--\eqref{e:nodim4} from $\xi=\pm\infty$ to 0 and
then find $\sigma$ by the requirement $\mathbf{U}=\mathbf{D}$ at
$\xi=0$. To do this, we need appropriate boundary conditions at
far-field zones as described below.

\subsection{State Vectors}

While $\mathcal{T}$ in our problem varies with $\xi$, there are
regions far away from the front where $\mathcal{T}$ can be treated
constant, thereby allowing algebraic solutions for perturbations.
More specifically, let $\xim\;(<0)$ and $\xip\;(>0)$ be the positions
in the upstream and downstream flow, respectively, such that
\begin{equation}\label{e:far}
  \left|\dfrac{d\ln\mathcal{T}}{d\xi}\right| \ll \text{min}(1,\nu),
\end{equation}
at the far-field zones with $\xi<\xim$ or $\xi>\xip$. The boundary
conditions for the perturbed variables in these regions are that they
should be regular as $\xi\rightarrow\pm\infty$, that is, the
perturbations should behave as $\propto e^{\beta\xi}$ for $|\xi|\gg
1$, with $\text{Re} (\beta) >0$ at $\xi<\xim$ and $\text{Re} (\beta) <
0$ at $\xi>\xip$.  Substituting the perturbations of this form into
the perturbed continuity and momentum equations (Eqs.\
\eqref{e:nodim1}--\eqref{e:nodim3}), one obtains
\begin{equation}\label{e:wave1}
  (\beta^2 - \nu^2)(\mathcal{T}\beta + \nu\sigma )\mathcal{J}^{\pr} =
  \dfrac{\nu}{\mathcal{T}^2}(\mathcal{T}\beta + \nu\sigma)(\sigma\beta
  + \nu\mathcal{T})\mathcal{T}^{\pr}\,,
\end{equation}
while Equation \eqref{e:nodim4} leads to
\begin{equation}\label{e:wave2}
  (\beta^2 - \beta + d\mathcal{H}/d\mathcal{T} - \nu^2 -
  \nu\sigma/\mathcal{T})\mathcal{T}^{\pr} = 0\,,
\end{equation}
in the far-field zones.

Clearly, there are five distinct values that $\beta$ can take. The
first three values can be obtained from Equation \eqref{e:wave1} by
imposing $\mathcal{T}^\pr=0$, corresponding to hydrodynamic waves
propagating from the front. These are $\beta_{i} =\pm \nu$
representing \emph{incompressible} modes,\footnote{In fact, $\beta^2 -
  \nu^2=0$ is the incompressible version of the more general
  dispersion relation $\beta^2=\nu^2 + (\sigma\beta +
  \nu\mathcal{T})^2 / a^{2}$ for acoustic modes, where $a\equiv
  (dP/d\rho)^{1/2}/v_{x1} \gg 1$ is the dimensionless sound speed.}
and $\beta_v = -\nu\sigma/\mathcal{T}$ representing a \emph{vortex}
mode carried by the background flow (e.g., \citealt{lan87}). The
remaining two solutions are
\begin{equation}\label{e:tmode}
  \beta_{t} = \frac{1}{2} \pm \sqrt{\frac{1}{4} + \nu^2 +
    \frac{\nu\sigma}{\mathcal{T}}-\frac{d\mathcal{H}}{d\mathcal{T}}}\,,
\end{equation}
obtained from Equation \eqref{e:wave2}, corresponding to
\emph{thermal} modes. The associated state vectors can be obtained by
substituting $\beta$'s back to Equations
\eqref{e:nodim1}--\eqref{e:nodim4}.

In general, perturbations that grow at far fields are a superposition
of these five basic modes, but the boundary conditions mentioned above
limit the number of the basic modes by requiring $\beta>0$ in the
upstream flow and $\beta<0$ in the downstream flow.  In addition, the
vortex mode is related to the advection of vorticity which is
generated when the front is curved. Since there is no source of
vorticity generation other than the front itself \citep{zel85}, the
vortex mode can exist only in the downstream flow if the flow at
$\xi=-\infty$ is irrotational.  Therefore, we are left with the
following five state vectors.
\begin{enumerate}
\item Upstream incompressible mode:
\begin{eqnarray}\label{e:v1}
  \mathbf{U}_{i} = e^{\nu\xi} (1,\, -1,\, -(\sigma - 1),\, 0,\, 0);
\end{eqnarray}
\item Downstream incompressible mode:
\begin{eqnarray}\label{e:v2}
  \mathbf{D}_{i} = e^{-\nu\xi} (1,\, \mu,\,
    \sigma + \mu,\, 0,\, 0);
\end{eqnarray}
\item Downstream vortex mode:
\begin{eqnarray}\label{e:v3}
  \mathbf{D}_{v} = e^{-\nu\sigma\xi/\mu} (1,\, \sigma,\,
    2\mu,\, 0,\, 0 );
\end{eqnarray}
\item Upstream thermal mode:
\begin{eqnarray}
  \mathbf{U}_t = e^{\betam\xi}
  \left(
    \nu \dfrac{\betam\sigma + \nu}{\betam^2 - \nu^2},\,
    -\nu\dfrac{\betam + \nu\sigma}{\betam^2 - \nu^2},\,
    \nu^2\dfrac{1 - \sigma^2}{\betam^2 - \nu^2},\,
    1,\, \betam \right);
\end{eqnarray}
\item Downstream thermal mode:
\begin{eqnarray}
  \mathbf{D}_t = e^{\betap\xi}
  \left(
    \dfrac{\nu}{\mu^2}\dfrac{\betap\sigma + \mu \nu}{\betap^2
      - \nu^2},\,
    -\dfrac{\nu}{\mu}\dfrac{\betap\mu + \nu\sigma}{\betap^2 - \nu^2},\,
    \dfrac{\nu^2}{\mu^2}\dfrac{\mu^2 - \sigma^2}{\betap^2 - \nu^2},\,
    1,\, \betap \right),
\end{eqnarray}
\end{enumerate}
where $\betam$ and $\betap$ denote the positive and negative values
of $\beta_t$, respectively, from Equation \eqref{e:tmode}. The
perturbations at far fields can then be written as
\begin{equation}\label{e:bd1}
  \mathbf{U}(\xi) = C_{1} \mathbf{U}_{i} + C_{2}
  \mathbf{U}_t,\;\;\text{for}\;\;\;\xi\leq \xim\,,
\end{equation}
\begin{equation}\label{e:bd2}
  \mathbf{D}(\xi) = C_{3} \mathbf{D}_{i} + C_{4}
  \mathbf{D}_{v} + C_{5} \mathbf{D}_{t},\;\;\text{for}\;\;\;\xi\geq
  \xip\,,
\end{equation}
where $C_i$'s are constants to be determined.

Unlike hydrodynamic waves for which $\beta \propto \nu$, thermal waves
always have $|\beta_t| \sim \text{max} (1,\nu)$, decaying on a length
scale shorter than the front thickness $\DLone$. The role of the
thermal modes thus becomes important when $\nu\simgt 1$, while it can
be ignored in the long wavelength limit. We note that the analysis
presented by \citet{ino06} for short-wavelength perturbations
considered only thermal modes as the basic states and ignored
hydrodynamic modes.

\subsection{Dispersion Relations}

Since Equations \eqref{e:nodim1}--\eqref{e:nodim4} are linear in the
perturbed variables, we may take $C_1=1$ without loss of generality.
Therefore the problem is reduced to finding the eigenvalue $\sigma$
and four proportionality constants $C_2, \cdots, C_5$ subject to five
constraints, $\mathbf{D}=\mathbf{U}$ at $\xi=0$.

\subsubsection{Long-wavelength Limit}

Before explaining the computation method for obtaining dispersion
relations for general $\nu$, we revisit the case of long-wavelength
perturbations with $\nu\ll 1$, for which the thermal front can be
treated as a discontinuous surface at $\xi=0$.  In this case, the
temperature distribution of the background flow can be taken as
$\mathcal{T}=1$ for $\xi<0$ and $\mathcal{T}=\mu$ for $\xi>0$, and the
upstream and downstream far-field zones extend to $\xim=0^-$ and
$\xip=0^+$, respectively.

It can be shown that the terms in the left-hand side of Equation
\eqref{e:nodim4} is of zeroth order in $\nu$, while the terms in the
right-hand side are of higher order (e.g., \citealt{lib08}).  Using
the equilibrium condition (Eq.\ [\ref{e:stn}]), one can show that the
solution of the zeroth-order terms in Equation \eqref{e:nodim4} is
given by
\begin{equation}\label{e:zero1}
\mathcal{T}^\pr = \xi_{T} \frac{d\mathcal{T}}{d\xi}\,,
\end{equation}
where $\xi_T$ is a small constant representing a shift of the front
in $\xi$ \citep{lib08}. This indicates that the thermal modes are
absent except near the discontinuous front (i.e., $C_2=C_5=0$).

By integrating Equations \eqref{e:nodim1}--\eqref{e:nodim3} across the
front and by keeping the first-order terms in $\nu$, one obtains
\begin{equation}\label{e:zero2}
  \mathcal{J}^\pr_+ = \frac{1}{\mu} \mathcal{J}^\pr_-\,,\;\;\;
  \mathcal{M}^\pr_+ = \mathcal{M}^\pr_-\,, \;\;\; %
  \mathcal{V}^\pr_+ = \frac{\mu-1}{\sigma} \mathcal{V}^\pr_-\,,
\end{equation}
where the subscripts ``$-$'' and ``$+$'' indicate the values evaluated
at $\xi=\xim$ and $\xi=\xip$, respectively. Inserting Equation
\eqref{e:zero2} into Equations \eqref{e:bd1} and \eqref{e:bd2} and
using the hydrodynamic state vectors given in Equations
\eqref{e:v1}--\eqref{e:v3}, one obtains the quadratic equation
\begin{equation}
  \sigma^2 + \frac{2\mu}{\mu+1} \sigma - \frac{\mu(\mu-1)}{\mu+1} = 0,
\end{equation}
for $\sigma$, the positive (unstable) solution of which is identical
to Equation \eqref{e:Omega0}.

\subsubsection{General Cases}

To obtain $\sigma$ for arbitrary $\nu$, we proceed by taking $\xim$
and $\xip$ sufficiently large to satisfy Equation \eqref{e:far} for a
background configuration at given $\Peq$. We then choose five trial
values for $\sigma$ as well as $C_2,\cdots,C_5$, and integrate
Equations \eqref{e:nodim1}--\eqref{e:nodim4} from $\xi=\xi_\pm$ to
$\xi=0$ to find $\mathbf{D}(0)$ and $\mathbf{U}(0)$, respectively.  We
then check if the two vectors connect smoothly at $\xi=0$. If the
relative difference $|\mathbf{D}(0)/\mathbf{U}(0) -1|$ is larger than
the tolerance limit (say, $\sim10^{-3}$), we change $\sigma$ and
$C_2,\cdots, C_5$ iteratively based on the Newton-Rhapson technique
until the smoothly-connecting solutions are obtained. We repeat the
calculations by varying $\nu$ to find a dispersion relation for given
$\Peq$. Figure \ref{f:gr} plots as the solid lines the growth rate
$\sigma$ for $\Peq/\kB=1700$, 1800, and $1900\punit$. Equation
\eqref{e:Omega0} is compared as the dotted lines, which agree very
well with the true dispersion relations at $\nu\ll1$. Note that the
DLI is stabilized by thermal conduction at $\nu>\nu_{\rm crit} \sim
(0.52-0.54)$.


\end{document}